\documentclass[aps,twocolumn,floatfix ,groupedaddress,nofootinbib]{revtex4-2}
\usepackage{csquotes}
\usepackage[table,xcdraw]{xcolor}
\usepackage{graphicx,color}
\usepackage{rotating}
\usepackage{float}
\usepackage{lipsum}
\usepackage[caption=false]{subfig}
\usepackage{amsmath}
\usepackage{amssymb}
\usepackage{multirow}
\usepackage{dsfont}
\usepackage[finalnew]{trackchanges} %for better track changes. finalnew option will compile document with changes incorporated.
\expandafter\let\csname equation*\endcsname\relax
\expandafter\let\csname endequation*\endcsname\relax
\renewcommand{\figurename}{Fig.}

\begin{document}

\title{Revealing an Unexpectedly Low Electron Injection Threshold via Reinforced Shock Acceleration}

\author{Savvas Raptis$^{1,*}$} 
\author{Ahmad Lalti$^{2,3}$, Martin Lindberg$^{4,5}$, Drew L. Turner$^{1}$, Damiano Caprioli$^{6}$, James L. Burch$^{7}$}

\affiliation{$^1$ The Johns Hopkins University Applied Physics Laboratory, Laurel, MD, USA \\
$^2$ Swedish Institute of Space Physics, Uppsala, Sweden \\
$^3$ Uppsala University, Uppsala, Sweden \\
$^4$ Division of Space and Plasma Physics — KTH Royal Institute of Technology, Stockholm, Sweden\\
$^5$ Department of Physics and Astronomy, Queen Mary University of London, Mile End Road, London E1 4NS, UK \\
$^6$ Department of Astronomy \& Astrophysics and E. Fermi Institute, The University of Chicago, 5640 S Ellis Ave, Chicago, IL 60637, USA \\
$^{7}$Southwest Research Institute, San Antonio, TX, USA}
\maketitle

$^*$ Corresponding author : savvas.raptis@jhuapl.edu
 \subsection*{Abstract}
Collisionless shock waves, found in supernova remnants, interstellar, stellar, and planetary environments, and laboratories, are one of nature's most powerful particle accelerators. This study combines \textit{in-situ} satellite measurements with recent theoretical developments to establish a novel reinforced shock acceleration model for relativistic electrons. Our model incorporates transient structures, wave-particle interactions, and variable stellar wind conditions, operating collectively in a multiscale set of processes. \change{Through these processes, we show}{We show} that the electron injection threshold is on the order of suprathermal range, obtainable through multiple different phenomena abundant in various plasma environments. Our analysis demonstrates that a typical shock can consistently accelerate electrons into very high (relativistic) energy ranges, refining our comprehension of shock acceleration while providing new insight on the origin of electron cosmic rays.
 
\subsection*{Introduction}

Understanding how charged particles in the universe gain immense  energy, reaching relativistic levels, remains a fundamental challenge in modern physics.  
Collisionless shocks are ubiquitously found throughout the universe (e.g., heliospheric \cite{chen2015particle}, astrophysical \cite{drury83,blasi13}, planetary \cite{masters2013electron},  and laboratory \cite{fiuza2020electron}) and are considered to be the primary drivers of accelerating charged particles to such an energetic regime \cite{caprioli2014simulations}. 
Therefore, they have been the main focus of research for understanding the cosmic ray (ultra-relativistic particles) profiles observed at Earth. Shock Drift Acceleration (SDA) occurs when a charged particle encounters a collisionless shock. This mechanism can typically accelerate electrons to relatively high energies, but it is not sufficient to explain the observations of relativistic electrons \cite{ball2001shock}. Diffusive Shock Acceleration (DSA), known as the first order Fermi mechanism, can accelerate particles to relativistic energies \cite{blandford1978particle}. However, to get particles efficiently accelerated through DSA, they need to be pre-energized to a certain threshold before they can be 'injected' into the acceleration process. Ion injection to DSA is sufficiently understood since ions have a large dynamical scale length, comparable to typical shock scales. On the other hand, electrons due to their much smaller gyroradius, have a higher injection threshold, requiring them to reach mildly relativistic energies ($10-100$ keV) before they can get further accelerated through DSA. Finding which process consistently allows these energies to be obtained in a collisionless shock is called the electron injection problem and remains one of the open questions of modern physics \cite{park2015simultaneous,matsumoto2015stochastic,shalaby2022mechanism}. In this work, we propose a potential resolution to this problem by showing the first \textit{in-situ} observations  of a new comprehensive multistep process operating across various scales, allowing electrons to consistently reach relativistic energies. As we present below, our model refines the traditional shock acceleration processes by introducing new findings revealing a new electron injection threshold. This result is obtained through encompassing recent advancements on astrophysical plasmas and \remove{nonlinear} wave-particle interactions, and is generalized to other stellar and interstellar environments.

The dynamics of a collisionless shock depend greatly on a key property, the angle between the shock normal vector and the magnetic field in the upstream plasma ($\theta_{Bn}$) \cite{caprioli2014simulationsb}. In a typical plasma consisting of electrons and protons, when this angle is approximately less than 45 degrees, \change{particles}{protons} (and heavy ions) can get reflected, travel far upstream, interact with the incident plasma, and cause a series of instabilities and wave activity forming an extensive region called the foreshock or precursor. The foreshock region consists of a strongly variable and dynamically evolving  plasma, while characterized by temporally varying (transient) and spatially localized structures. Subsequently, this extends the shock’s presence and intensifies its impact on the nearby plasma. As a result, in our Solar System, quasi-parallel shocks ($\theta_{Bn} \le 45^\circ$) typically emerge as notably more efficient particle accelerators compared to quasi-perpendicular ($\theta_{Bn} \ge 45^\circ$) shocks \cite{masters2013electron,turner2018autogenous,liu2019relativistic}. Foreshock transients \add{(or dayside/shock-generated transients)} occur multiple times per day \cite{schwartz2000conditions} and form when the solar wind and its embedded, variable and discontinuous magnetic field interacts with a shock wave, like Earth's bow shock \add{\protect{\cite{schwartz2000conditions,zhang2022dayside}}}. Foreshock transients are also recognized for their ability to accelerate particles to high energies \cite{wilson2016relativistic,turner2018autogenous,liu2019relativistic}.  \add{The transients examined in this study are commonly referred to as Hot Flow Anomalies (HFAs) in heliophysics literature. However, due to their similarities with other transients, such as Foreshock Bubbles (FBs), which also play a role in particle acceleration, we use the more general term 'foreshock transient' throughout the text.}

\subsection*{Results}

The model presented in this work primarily relies on \textit{in-situ} observations from two NASA missions. The Magnetospheric Multiscale (MMS) mission \cite{burch2016magnetospheric} provides near-Earth observations, while the Acceleration, Reconnection, Turbulence, and Electrodynamics of Moon's Interaction (ARTEMIS) mission \cite{angelopoulos2014artemis} describes the far upstream plasma environment close to the dayside Moon. On December 17, 2017, between 17:50 UT and 17:55 UT, the MMS spacecraft were positioned just upstream of Earth's bow shock and observed an unprecedented event. MMS recorded a foreshock transient associated with the highest energetic electrons ever observed upstream of the bow shock since its prime mission started in 2015, under relatively steady solar wind conditions. Electrons reached observable intensity enhancements to more than 500 keV, a remarkable feat considering the typical observable range is up to only a few keV in the solar wind near Earth. This is even more extraordinary as there was an absence of significant solar disturbances, like a flare or a coronal mass ejection (CME), which could otherwise explain the presence of such relativistic electrons. Our findings indicate that the electron acceleration during this event manifested within a foreshock transient. This phenomenon emerged due to a disturbance (discontinuity) in the interplanetary magnetic field, interacting with Earth's bow shock and foreshock plasma. ARTEMIS, positioned in lunar orbit over one hundred thousand kilometers towards the Sun from MMS, observed this magnetic disturbance along with an associated electron seed population a few minutes prior to the foreshock transient formation.

\begin{figure*}[ht]
    \centering
{\includegraphics[width=1\textwidth]{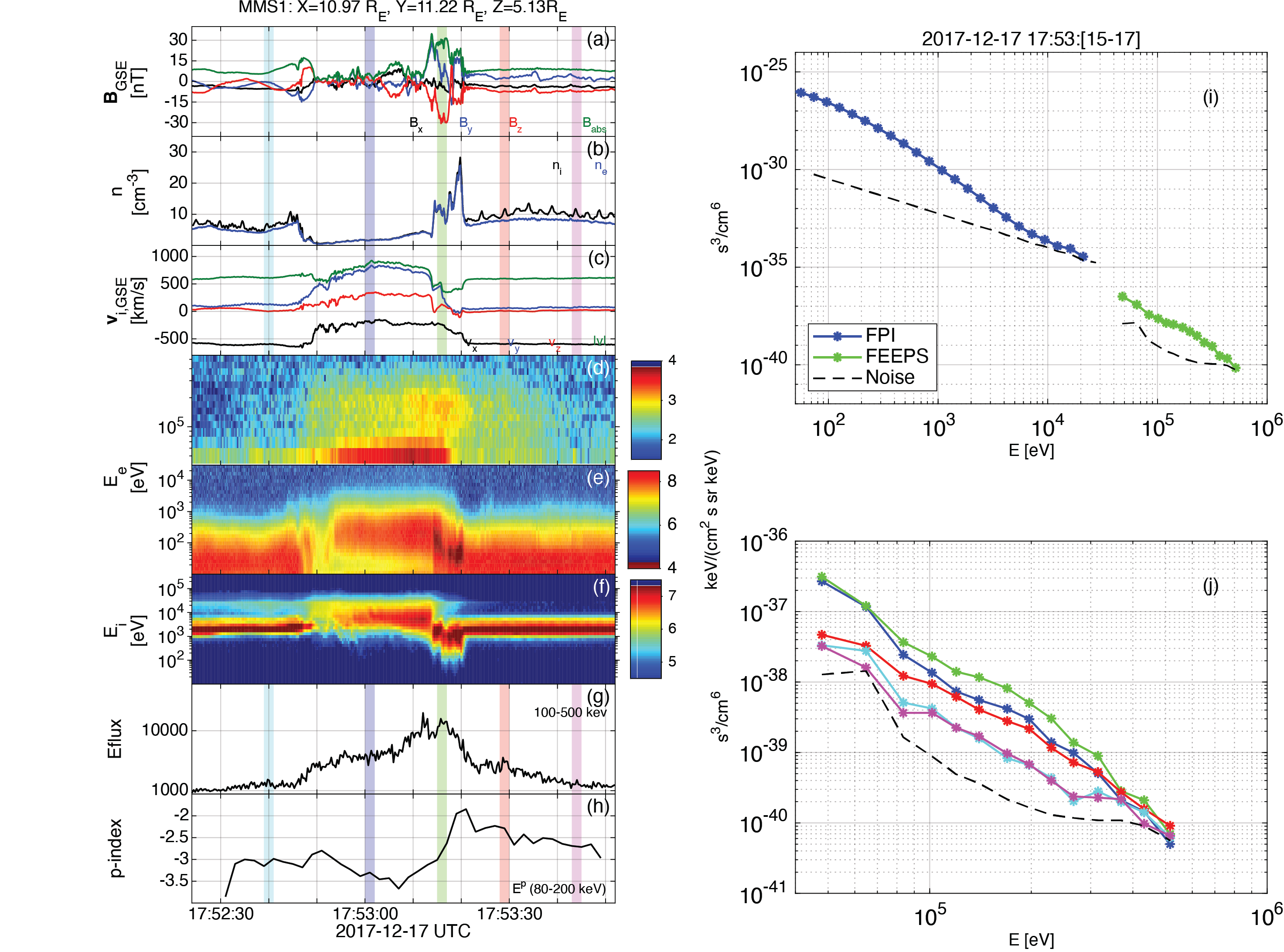}}\caption{
\textbf{Overview observations of the main electron acceleration event}. (a) magnetic field components and magnitude in Geocentric Solar Ecliptic (GSE) system, (b) ion and electron densities, (c) ion plasma velocity components and magnitude in GSE coordinates, (d) high-energy (FEEPS) differential energy flux electron spectra, (e) low-mid energy (FPI) differential energy flux electron spectra, (f) combined (FPI + FEEPS) differential energy flux ion spectra, (g) integrated energy electron flux between 100 and 500 keV, (h) fitted spectra index for $E^p$ between 80-200 keV, (i) electron phase space density (PSD) versus energy for the whole energy range (FPI + FEEPS) along with background noise level, (j) series of PSD versus energy line plots for high-energy measurements over different time intervals. The colors of panel (j) are indicated as highlighted areas in the time series plot of panels (a-h). The event is observed from 17:52:45 to 17:53:25. Panels (d) and (g) show a distinct enhancement of relativistic energies in the 100-500\,keV range. The green shaded area around 17:53:15-20 shows the area that has the strongest acceleration. The electron enhancement there can be seen compared to other times by looking at panel (j). The title shows the location of MMS spacecraft in GSE coordinates given in Earth radius units.  More information regarding the derivation of each product can be found in the Methods section}
\end{figure*}

\begin{figure*}[ht]
    \centering
{\includegraphics[width=1\textwidth]{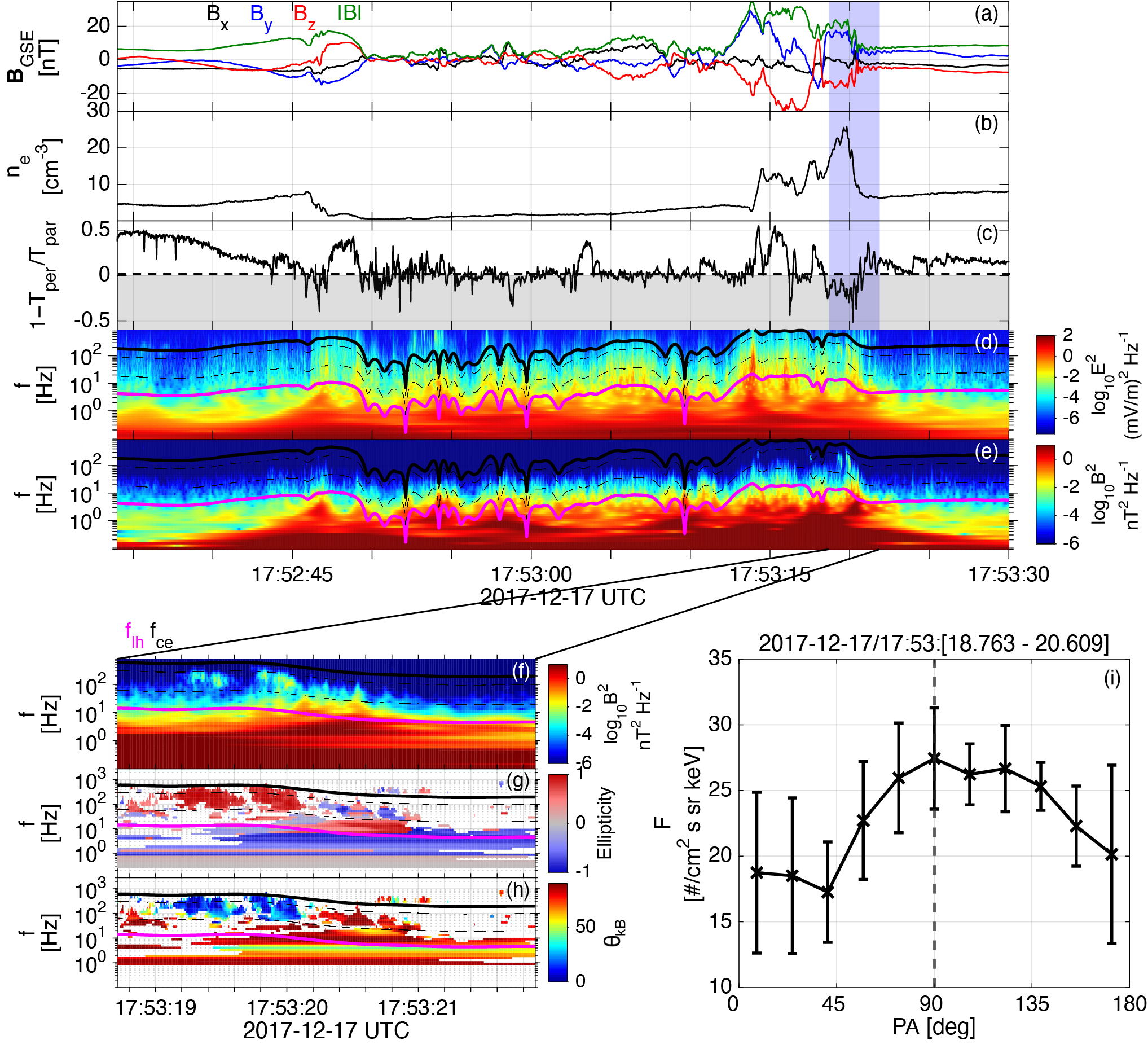}}\caption{
\textbf{Detailed plot of the main electron acceleration event and wave analysis}. (a) magnetic field components and magnitude in GSE coordinates, (b) electron density, (c) electron temperature anisotropy, for which gray shaded are indicates the region where negative temperature anisotropy takes place, (d) electric field power spectra, (e) magnetic field power spectra, (f) zoomed-in plot of magnetic field power spectra, (g) ellipticity showing the polarization of the magnetic field power spectra, (h)  wave propagation angle, (i) pitch angle distribution (PAD) for electrons between 40 and 200 keV from the FEEPS instrument averaged over 6 measurements when the wave activity is the strongest as indicated by the date at the title \add{Error bars represent the Standard Error (SE) for each measurement}.  Nonlinear electron whistler waves are present in the large-amplitude, banded emissions between 0.1 and 1.0 of the electron cyclotron frequency, $f_{ce}$, with ellipticity near 1 (red in panel g) and propagation angles parallel to the background magnetic field (blue in panel h). 0.1 and 1.0 $f_{ce}$ are shown with the bottom-most dotted and solid black lines in the wave panels; the magenta line and the middle dashed line in the wave panels are the plasma lower hybrid frequency and 0.5 $f_{ce}$, respectively. Such nonlinear whistler-mode wave packets are capable of resonantly interacting with relativistic electrons, resulting in predominantly perpendicular acceleration (as shown in panel i). More information regarding the derivation of each product can be found in the Methods section.}
\end{figure*}

\begin{figure*}[ht]
    \centering
{\includegraphics[width=0.75\textwidth]{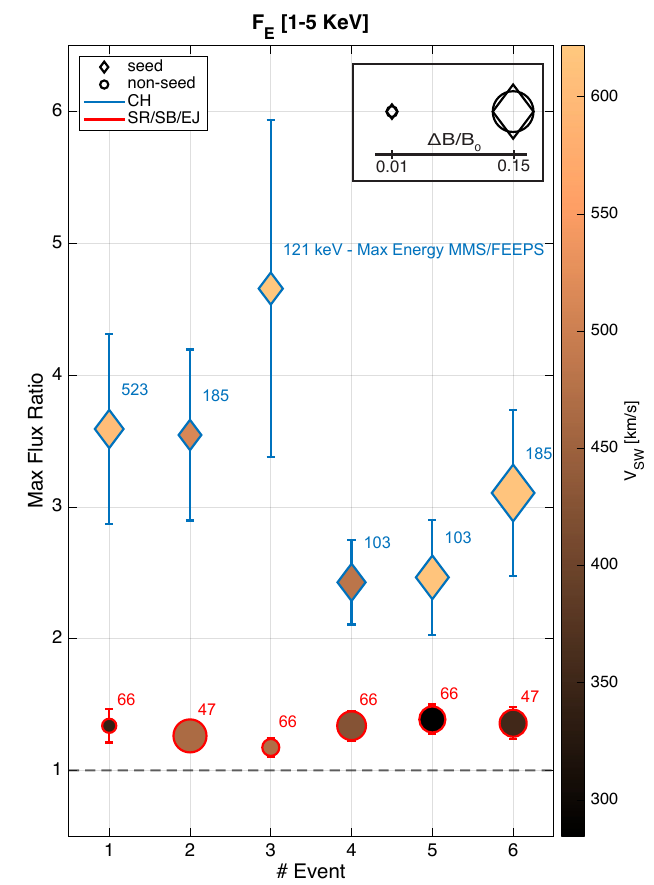}}\caption{
\textbf{Statistical analysis and validation of the reinforced shock acceleration for relativistic electrons}. The analysis includes six events with and without the presence of different components that embody our model. The y-axis shows the maximum ratio between the 1-5 keV flux of the seed population with respect to the background from the initial solar/stellar wind, while the x-axis indicates the number of each event. The colorbar shows the bulk flow velocity of the stellar wind, revealing the connection between the high-speed stream and the higher maximum electron flux ratio. The marker color shows if the particles originate from fast stellar wind (blue) or slow stellar wind (red). CH refers to coronal  hole plasma, SR to sector reversal, SB to streamer belts and EJ to ejecta (\cite{xu2015new,borovsky2016plasma,camporeale2017classification}). The marker shape shows the presence of the initial seed enhancement, with a diamond indicating the seeded events and with a circle for the non-seeded ones. The different marker size indicates the amplitude of the nonlinear electromagnetic waves compared to the background ($\delta B/B_0$), while each event is accompanied by text referring to the energy channel for which a significant electron flux was measured above the noise instrument level close to Earth. More information regarding the derivation of each product and the classification of the solar wind can be found in the Methods section. The exact times for which we took measurements for MMS and ARTEMIS are given in Methods section at Table. \ref{tab:S4}. \add{Error bars represent the Standard Error (SE), calculated using the information described in the Methods section.}}
\end{figure*}

\begin{figure*}[ht]
    \centering
{\includegraphics[width=0.47\textwidth]{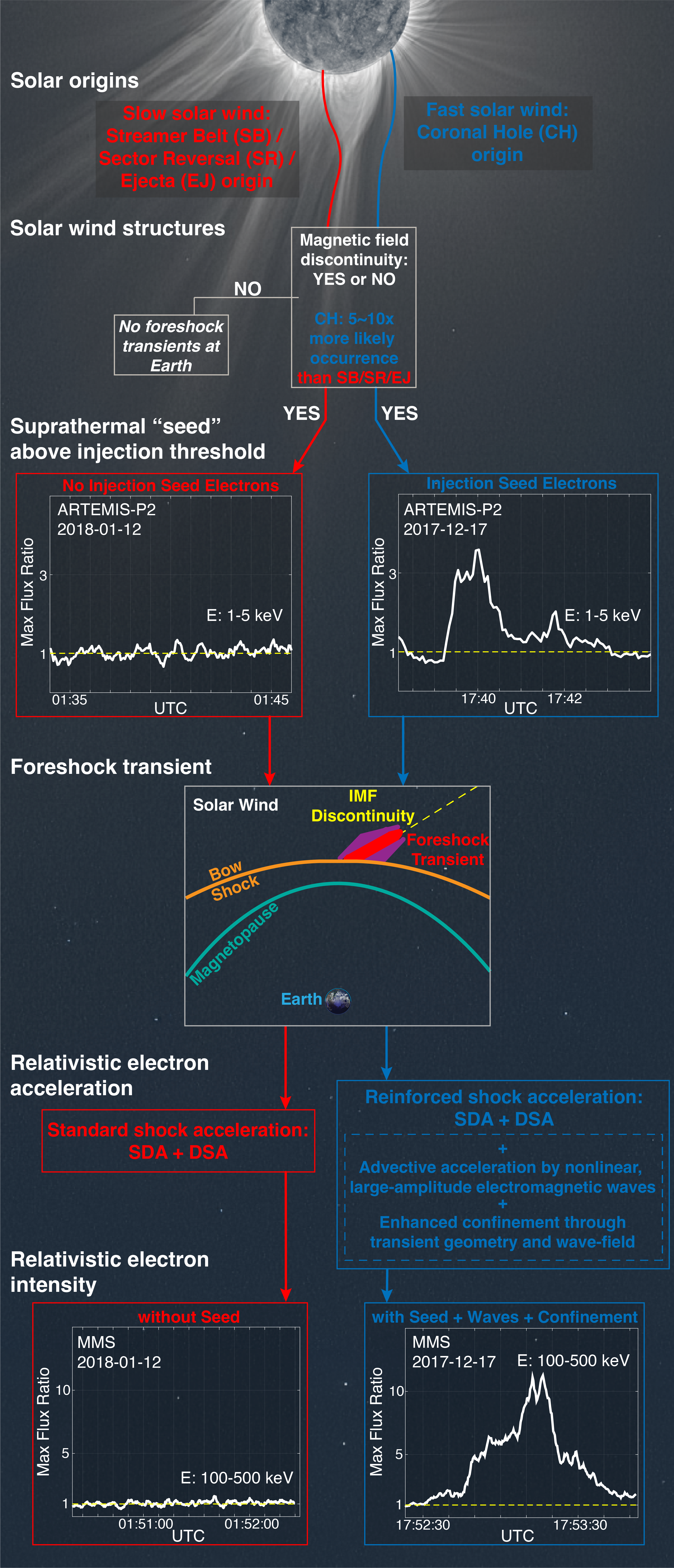}}\caption{
\textbf{Schematic of the reinforced shock acceleration mechanism}. The process unfolds in several steps, illuminating the distinctions between two distinct events: the primary seeded event marked in blue and a non-seeded event in red. The first step is the fueling of the process, originating directly from the Sun's activity, in which the fast solar wind plasma provides a seed population above the electron injection threshold (1-5\,keV) associated with magnetic field discontinuities as measured by the ARTEMIS mission during its dayside Lunar orbit. The next step involves a Fermi-type shock (i.e., DSA) \add{and/or stochastic shock drift (SDA)} acceleration, occurring at the sunward-facing shock within a foreshock transient. This transient arises after the discontinuity interacts with Earth's bow shock. The acceleration of electrons is further enhanced by the presence of \change{nonlinear}{wave-particle} interactions between electrons and high-amplitude, high-frequency electromagnetic waves. Lastly, the process's efficiency is intensified by particle scattering and confinement within the local magnetic topology of the transient and its connectivity to the primary shock (i.e., Earth’s bow shock in this case). Electromagnetic waves within the foreshock transient effectively scatter particles back to the acceleration region, while the surrounding shock's geometry acts as a confining boundary, enabling electrons to accumulate and obtain relativistic energies before reaching Earth and being lost to the downstream medium. The image of the Sun was taken during a solar eclipse and credits are attributed to Miloslav Druckmüller and Peter Aniol (http://www.zam.fme.vutbr.cz/~druck/Eclipse/Ecl2006l/0-info.htm)}\label{fig:4}
\end{figure*}

In Fig. 1, we present an overview plot highlighting the observations, emphasizing the presence of high-energy electrons above the rest mass energy ($>$511 keV) as shown in panels (d) and (g). As displayed in panels (a-h) around the green shaded area, a compressive edge resulting in a fast collisionless quasi-perpendicular shock is formed, which is a typical feature of foreshock transients \cite{liu2019relativistic,turner2021direct}. Another vital feature of the foreshock transient is shown around the purple shaded area, where the “core” of the structure is infused by high-amplitude electromagnetic waves  over a wide range of frequencies \add{(illustrated in Fig. 2)}, in a low-density plasma environment. This wave and topological environment facilitates electron scattering, enabling them to cross the shock (green shaded area) multiple times, where typical shock acceleration mechanisms occur (i.e., DSA/SDA) \cite{park2015simultaneous,amano2020observational}. 
However, in contrast to theoretical and observational expectations of quasi-perpendicular SDA \cite{amano2020observational}, in panel (h), the energy power-law spectral index reaches canonical values up to $p \sim -2$, which is the theoretical \change{limit}{prediction} for DSA at strong shocks \add{\protect\cite{axford1977acceleration,longair2011high,oka2018electron}}. \add{
This indicates that the acceleration observed upstream of the planetary bow shock results in an even harder spectrum than what would typically be produced by a classical DSA mechanism.} One direct explanation for this observation is that in foreshock transients, the wave field is found in a lower background magnetic field than a typical downstream shocked plasma, which allows particles to get more efficiently scattered and subsequent acceleration to take place \cite{liu2017fermi,shi2020whistler}. Moreover, these electrons can exhibit a bouncing behavior between Earth's nearby shock environment and the foreshock transient. This interplay results in ideal confinement, significantly enhancing the effectiveness of acceleration compared to a typical quasi-perpendicular shock \cite{liu2017fermi,turner2018autogenous,shi2020whistler} \add{while other factors such as the presence of electrostatic waves can also play a role in the scattering of particles near the shock transition \protect{\cite{vasko2018solitary,kamaletdinov2020superthin,kamaletdinov2024nonlinear}}}.

In Fig. 2, a crucial factor reinforcing the shock acceleration process is unveiled. Specifically, by analyzing the compressive (shock) region of the foreshock transient, the interaction between high-frequency, high-amplitude, banded emissions of nonlinear electromagnetic waves and the local electron populations is shown. Panel (c) of Fig. 2 displays a notable electron temperature anisotropy ($A = 1 - T_{per}/T_{par}<0$) known to drive large-amplitude, high-frequency electromagnetic waves, termed whistlers (or “chorus” in magnetospheric research nomenclature), which can experience instability and wave growth from electron betatron acceleration localized within the compression region \cite{liu2019relativistic,shi2023intense}. The presence and characteristics of these waves enables higher energy electrons to cyclotron resonate with them, amplifying the electrons' energy \remove{and driving them to relativistic levels} through \remove{nonlinear (not to be confused with quasi-linear wave theory)} wave-particle interactions \cite{bortnik2008nonlinear}, thereby further enhancing the shock acceleration process \cite{thorne2013rapid}. This phenomenon has been observed in magnetospheric environments \cite{horne2003resonant,thorne2013rapid} and recently in collisionless shocks \cite{shi2023intense}. The nature of these waves is identified by the characteristic narrowband frequency in the magnetic field power spectra between 0.1 and 1.0 of the local electron cyclotron frequency (panels e and f), the right-hand polarization (panel g), and the propagation parallel to the magnetic field (panel h). Finally, cyclotron-resonant acceleration of relativistic electrons is expected to result in a predominantly perpendicular (with respect to the background magnetic field) acceleration \cite{horne2003resonant,thorne2013rapid}. As shown in panel (h), the pitch angle distribution (PAD) confirms that theoretical expectation. The electrons' pitch angle distribution peaks at 90 degrees with respect to the local magnetic field as soon as the spacecraft crosses the wave acceleration region, in \remove{full} agreement with the wave-particle acceleration model. \add{It should be noted that other mechanisms, such as betatron acceleration, could produce similar pitch angle distributions (PADs) and are expected to contribute to the observed acceleration. However, evaluating the contribution of each mechanism is not feasible observationally and falls outside the scope of this study.}

At this point, we have confirmed  the presence of multiple acceleration and confinement components taking place from small (i.e., plasma kinetic) to large (global with respect to Earth’s bow shock within the solar wind) scales. Specifically, \add{we have shown }the presence of shock acceleration in the foreshock transient \cite{turner2018autogenous,liu2019relativistic}, a strong wavefield and confining region \cite{liu2017fermi,shi2020whistler}, and nonlinear wave-particle interactions and acceleration between the electromagnetic whistler waves and the electron distribution \cite{artemyev2022electron,shi2023intense}. Furthermore, the environment between the primary planetary bow shock and the foreshock transient creates a larger bounded environment in which the particles can remain confined, bouncing between the regions \cite{liu2017fermi,turner2018autogenous}, this is particularly true for magnetic field discontinuities that intersect the bow shock under certain geometries like in our case (see Methods section). However, these elements cannot entirely explain the distinctive presence of electrons with energies \change{$\ge 0.5 $ MeV}{$\sim$ 500 keV}, nor the occasional absence of such particle population during foreshock transients with similar characteristics \cite{liu2017fermi}. Hence, we further explored the possibility of identifying whether a specific seed population, well beyond the vicinity of Earth's bow shock environment, was present, corresponding to global system size effects. From a multi-case study described below, our investigation revealed a clear increase in the suprathermal electron flux (1-5 keV) measured by the ARTEMIS spacecraft, in dayside lunar orbit, during all the foreshock transient events associated with significant electron acceleration, as observed by MMS. Additionally, this seed electron population was associated with faster-than-average solar wind, typical of what emerges from solar coronal holes at higher solar latitudes, resulting in high-speed plasma streams \cite{feldman2005sources,xu2015new,borovsky2016plasma}.

To validate our model, we extensively reviewed all available observations from the operational years of MMS. While observational limitations were present (MMS required to be upstream of the shock with burst data availability and ARTEMIS in dayside far-upstream lunar orbit), we found several events aligning, either partially or fully, with our model components. Upon analysis, we noticed a distinct pattern. Whenever MMS recorded electron fluxes exceeding 100 keV upstream of the bow shock, ARTEMIS detected a clear seed population of suprathermal, solar wind electrons. This trend consistently occurred when the solar wind originated from the Sun's coronal holes, typically referred to as fast solar wind  \cite{feldman2005sources,xu2015new,borovsky2016plasma,camporeale2017classification}. Our findings were further supported by the increased occurrence of magnetic field discontinuities and overall variability (necessary ingredients for the formation of foreshock transients) in fast solar wind plasma \cite{tsurutani1999review,liu2022magnetic}. In all events associated with fast coronal hole plasma, the suprathermal electron flux (1-5 keV) was notably elevated, ranging from 2 to 5 times higher than the background level (see Fig. 3). In contrast, slow solar wind events in which no relativistic electron acceleration occurred were lacking an enhanced seed population. As a result, despite the presence of magnetic discontinuities and consequently the formation of foreshock transients, MMS measured primarily background noise on the high-energy particle instruments. The source of the initial seed population could potentially arise from magnetic reconnection in the solar wind \cite{gosling2005direct}, turbulence \cite{breech2009electron}, discontinuities impacting the solar wind plasma \cite{shen_comparing_2024} or direct jet outflow from the coronal hole environment \cite{raouafi2023magnetic}. Furthermore, by using the high-energy instrument on board of ARTEMIS and removing events flagged by low quality, we confirm that relativistic electrons were not present in lunar orbit for the main event analyzed and for most of the seeded events. This allowed us to rule out the possibilities of pre-accelerated relativistic electrons coming from external origin. Finally, the enhanced suprathermal electron flux at ARTEMIS contain both an earthward and sunward components, while the magnetic connectivity to the bow shock is taking place only for a couple of cases. This indicates that, a significant portion of the observed seed population appears to be of Solar origin. However, the connectivity to the shock and the presence of sunward particles in some cases hints towards an interplay between the solar wind and the Earth's bow shock under which particles from the electron foreshock may contribute to the seeding of the acceleration mechanism (see e.g., \cite{liu2017fermi}). More details are shown in methods - \textit{Statistical analysis details}. \add{Finally, we should note that while the presence of fast coronal hole (CH) solar wind and associated suprathermal energy elevation is a necessary condition for the acceleration of high-energy electrons, it is not sufficient on its own. There are instances where, despite the fast solar wind conditions, the acceleration mechanisms are limited, resulting in an absence of high-energy (100+ keV) electrons.}

To complete our analysis, we investigated whether the energies observed at the most energetic of our events (Fig. 1 and 2) \change{is}{are} bounded by diffusion scale lengths of Earth's bow shock environment. For plasma environments with wavefields corresponding to significantly high nonlinearity (i.e., $\delta B \sim B_0$) like the foreshock transient's core region (purple shaded area Fig. 1), the diffusion length is approximated by assuming Bohm diffusion \cite{caprioli2014simulations}. This means that the effective mean free path of electrons in these environments is approximately equal to the electron gyroradius (i.e., $\lambda \sim r_{ce}$) \cite{caprioli2014simulations}. Using that, we recovered that for the maximum energy observed ($\sim0.5$ MeV) we obtain, a diffusion length to be in the order of 10-100 $\mathrm{R_e}\approx 6\cdot 10^{4-5}$ km \add{(see more details in methods - \textit{Diffusion time scales and spatial scale sizes)}} which is on the same order as foreshock transients' scale sizes observed at Earth \cite{valek2017hot}.  

This finding is particularly interesting, as other planetary environments with larger system scales sizes (e.g., Jupiter) can sustain higher obtainable maximum energy. By using a fixed diffusion length and magnetic field, we can  calculate the highest energy that an electron can obtain before the scale of their gyroradius is too large for it to remain in the shock acceleration region. Specifically, by using a diffusion length of $L\sim100-1000 \; \mathrm{R_e} \approx  6\cdot10^{5-6}$ km and a background magnetic field of $1$ nT, we can obtain higher relativistic energies up to $E\sim2$ MeV. which is in agreement with previous observations and analysis \cite{masters2009hot,valek2017hot}. It should be noted that this upper energy can be significantly higher in other astrophysical systems (e.g., ultrahot Jupiters \cite{cauley2019magnetic}). Following the same argumentation as above, our mechanism shows that such a system can sustain electrons in the order of GeV to low TeV range (even when considering synchrotron losses), corresponding to the highest, ultra-relativistic energies of cosmic ray electrons observed at Earth \cite{nichols2016stellar,chebly2023numerical}.

Finally, in Fig. 4, we provide an illustrative summary of the proposed, reinforced shock acceleration model, depicting the evolutionary pathway of two distinct events. The first event, showcased in blue on the right-hand side of the schematic, corresponds to the main seeded event detailed in Fig. 1 and 2 (also referenced as \#1 seeded event in Fig. 3). In contrast, the left-hand side of the schematic portrays a non-seeded event (highlighted in red), identified as \#1 non-seeded in Fig. 3. This schematic comparison offers a clear visualization of the differences between each possible scenario highlighted by \textit{in-situ} measurements of both ARTEMIS and MMS mission.

\subsection*{Discussion \& Conclusion}

Our work engendered a unified shock acceleration mechanism applicable to stellar and interstellar plasma environments. First, we revealed a surprisingly low electron injection threshold, observed for the first time in space plasmas. This threshold at Earth system is found to be in the order of suprathermal range ($\sim1-5$ keV). These suprathermal electrons occur under fast solar wind conditions and are sufficient to enable a reinforced shock acceleration mechanism. The presence of this seed population, allows the foreshock transient to consistently produce relativistic electrons \cite{turner2013first}.  This is the result of a particularly efficient multiscale process (acceleration efficiency: $\mathrm{\nu = \frac{U_e}{U_i} \approx 0.05 = 5\%}$ \add{(see more details on methods section}). On ion kinetic scales, the foreshock transient, along with its compressive edge forms, allowing shock acceleration to occur ahead of the primary planetary bow shock \cite{wilson2016relativistic,liu2019relativistic}. On electron kinetic scales, within the foreshock’s shock, wave-particle energy transfer from the nonlinear high-frequency, high-amplitude electromagnetic waves contribute to the acceleration of suprathermal electron to relativistic energies \cite{artemyev2022electron,shi2023intense,li2024identification}. \add{This resonance mechanism occurs alongside larger-scale processes such as betatron and Fermi acceleration, which take place within the foreshock transient scales \protect{\cite{liu2017fermi,liu2019relativistic,zhang2022dayside}}}. Moving to \add{even} larger scales, as electrons emerge from the \add{foreshock transient's} shock acceleration region, they are confined by the presence of a highly nonlinear wavefield within the foreshock transient’s core and by the presence of the primary planetary bow shock that geometrically bound the electrons. We should note that since this whole process is taking place upstream of the primary shock, the already accelerated electrons can get additional energized via shock processes at the primary shock \cite{park2015simultaneous, oka2017electron,amano2020observational}.

 Undoubtedly, the most vital result we showed is that for Earth's shock system, the electron injection threshold in our model can be in the order of low $\sim$1-10 keV range (seed population in solar/stellar wind). This relatively low level of energy requirement in electron flux can be achieved through multiple mechanisms, yet it appears to be occurring exclusively under coronal hole solar wind plasma, corresponding to relatively faster than typical solar wind conditions.  As the solar activity is approaching its maximum (2024-2025), we expect solar focused missions such as Parker Solar Probe (PSP) and Solar Orbiter in conjunction with magnetospheric missions to shed light on the nature and origin of this seed population embedded within the high speed solar wind stream.

\add{Additionally, our results significantly enhance the understanding of electron acceleration by providing a clear explanation for past observations of relativistic electrons at Earth. Previously, these cases were puzzling due to their unusually high-energy populations. By re-evaluating the three events discussed in} \add{\protect{\citet{wilson2016relativistic}} and the event in \citet{liu2019relativistic}}, \add{we identify consistent conditions across all four events. Specifically, each event observed by THEMIS, like our findings, is associated with distinct fast coronal hole solar wind. In the few cases where ARTEMIS provided undisturbed upstream data, an elevated suprathermal electron flux was also observed. This insight is crucial, as it clarifies the unique nature of these foreshock transients, supporting our conclusions and providing the missing piece needed to explain previous results.}

Moving on, the generalization of this model is straightforward, since its ingredients are essentially fundamental astrophysical plasma processes (collisionless shocks and wave-particle interactions). The presence of the foreshock transients within our Solar System is also consistently found in all planetary systems where sufficient \textit{in-situ} measurements are available. Furthermore, the transients’ size scales with the planetary system size, allowing more efficient acceleration to occur in larger collisionless shock systems \cite{masters2009hot,collinson2014survey,collinson2015hot,valek2017hot}. Within our Solar System, the larger system size of gas giants (e.g., Jupiter) allows our mechanism to sustain energization of particles up to MeV range, consistent with previous observations. However, further evaluation by the planetary research community is crucial for the validation of our work and for the unification of Heliophysics subdomains. 

Apart from addressing electron acceleration within our Solar System, the implications of this model are directly influencing the origin of the electron cosmic ray profile. Cosmic Rays are thought to originate from high-mach number supernova shocks ($M_A > 10$) \cite{reynolds2011particle} although determining the exact details is an active research topic. However, it should be noted that even higher Mach number shock waves can be found in other planetary systems (e.g., Saturn \cite{masters2013electron} and Jupiter \cite{hospodarsky2017jovian}), while collisionless bow shock environments with similar intrinsic properties are found in the interstellar medium around young stellar systems \cite{ray2023outflows}. More importantly, in other stellar systems, under the presence of exoplanets like ultrahot Jupiters \cite{ cauley2019magnetic}, the existence of massive magnetic fields, enables our mechanism to potentially sustain GeV-TeV electrons. It should be noted that for young stellar systems the stellar wind can carry even stronger magnetic fields, that could sustain electrons to even higher energy ranges. Our results therefore imply that a significant portion of the cosmic ray distribution of relativistic electrons might originate from the interaction of planetary quasi-parallel shocks with typical stellar winds. This energy range is well within what is considered to be the ultra-relativistic electron range of cosmic rays. Further investigation should be made by the stellar astrophysics and particle acceleration communities to determine whether such unique systems (young stellar environments with ultra-hot Jupiters) can have a more significant contribution to the ultra-relativistic electron spectra through the utilization of the reinforced shock acceleration model. These stellar environments would therefore be particularly good candidates to investigate by looking for potential synchrotron emission. This implication originates directly from our model and its validity needs to be studied in the future, especially since up to today the main sources of these relativistic energies have been either pulsars or supernovas \cite{chang2008excess,archer2018measurement}. To fully address this impact, future endeavors will require coordinated cross-disciplinary efforts by the Heliophysics and Astrophysical communities, through the use of both remote sensing observations and computer simulations. \add{Future research in Earth's geospace environment could benefit from incorporating solar wind monitors at the Lagrangian 1 (L1) point alongside near-Earth observations from MMS and THEMIS. This approach would provide a more direct connection between coronal hole solar wind and the seed population. Additionally, large-scale global simulations that include nonlinear kinetic processes are needed to quantify the precise effects and relative contributions of each acceleration mechanism and scattering/confining process demonstrated in our observations.}

Finally, by reinforcing the acceleration processes at quasi-parallel shocks and by evaluating the electron injection threshold \change{of DSA}{for obtaining relativistic electrons at Earth} at suprathermal energy ranges, our model ultimately supports the role of quasi-parallel shocks as the source of the ultra-relativistic cosmic ray profile \cite{crumley2019kinetic}.

\clearpage

\subsection*{Methods}
\noindent\textbf{Data}. For Magnetospheric Multiscale (MMS) data \cite{burch2016magnetospheric}, we used burst resolution Level-2 data from the Fast Plasma Investigation (FPI), Flux Gate Magnetometer (FGM), Search Coil Magnetometer (SCM), Fly's Eye Energetic Particle Spectrometer (FEEPS). For the position of the spacecraft, we used the Magnetic Ephemeris Coordinates (MEC). For Acceleration, Reconnection, Turbulence, and Electrodynamics of Moon's Interaction (ARTEMIS) measurements \cite{angelopoulos2014artemis}, we used the Flux Gate Magnetometer (FGM), Solid Solid State Telescope (SST), and the Electrostatic Analyzer (ESA) and the state data for its position. Due to the close configuration of the MMS1-4 and ARTEMIS P1-P2, we primarily used MMS1 and ARTEMIS P2 for the analysis shown in the manuscript unless stated otherwise. Additionally, for validation purpose we used OMNIweb data to characterize the global scale of the solar wind conditions. The products used are the 1-h dataset and the propagated to the bow shock 1-min one, both accessible via \url{https://spdf.gsfc.nasa.gov/pub/data/omni/}.\\

\noindent\textbf{Coordinates system and spacecraft position}. Determining the connection between the MMS and ARTEMIS data required the use of position data for each spacecraft. We used the position of each satellite in Geocentric Solar Ecliptic (GSE) coordinates. In this system, the X-axis on the Sun-Earth line, while the Y-axis lies on the ecliptic plane towards the dusk. The Z-axis completes the coordinate system by being perpendicular to the ecliptic plane. The position of the spacecraft during the main event in GSE coordinate and in Earth Radius units ($\mathrm{R_E} \approx 6371$  km) is given as: MMS1: [10.8, 11.1, 5.1] and ARTEMIS P2: [65.2, -5.3, 5.1]. Upon using the solar wind velocity and convection along the X-GSE axis, we obtained an estimate \add{of the }time lag between the two spacecraft in the order of $\sim$10-15 mins.  The rest of the MMS satellite\add{s} (i.e., MMS2-4) are located in a close tetrahedron formation within a few kilometers from each other. This close formation allows multi-spacecraft methods such as timing to be used throughout our analysis, which is a typical method used in multi-spacecraft research \cite{turner2021direct,amano2020observational}. It should be noted that the position of ARTEMIS and MMS is very different on each event (see Table. \ref{tab:S4}). This results in a time lag between the spacecraft that can vary from instantaneous observations to tens of minutes depending on the orientation of the magnetic field discontinuity with respect to the plane MMS and ARTEMIS reside. Essentially, the 3D geometry of the magnetic field and the fact that discontinuities are planar structures allows any kind of time lag between the signals (see more details on the following subsections).

\begin{figure*}[ht]
    \centering
{\includegraphics[width=0.8\textwidth]{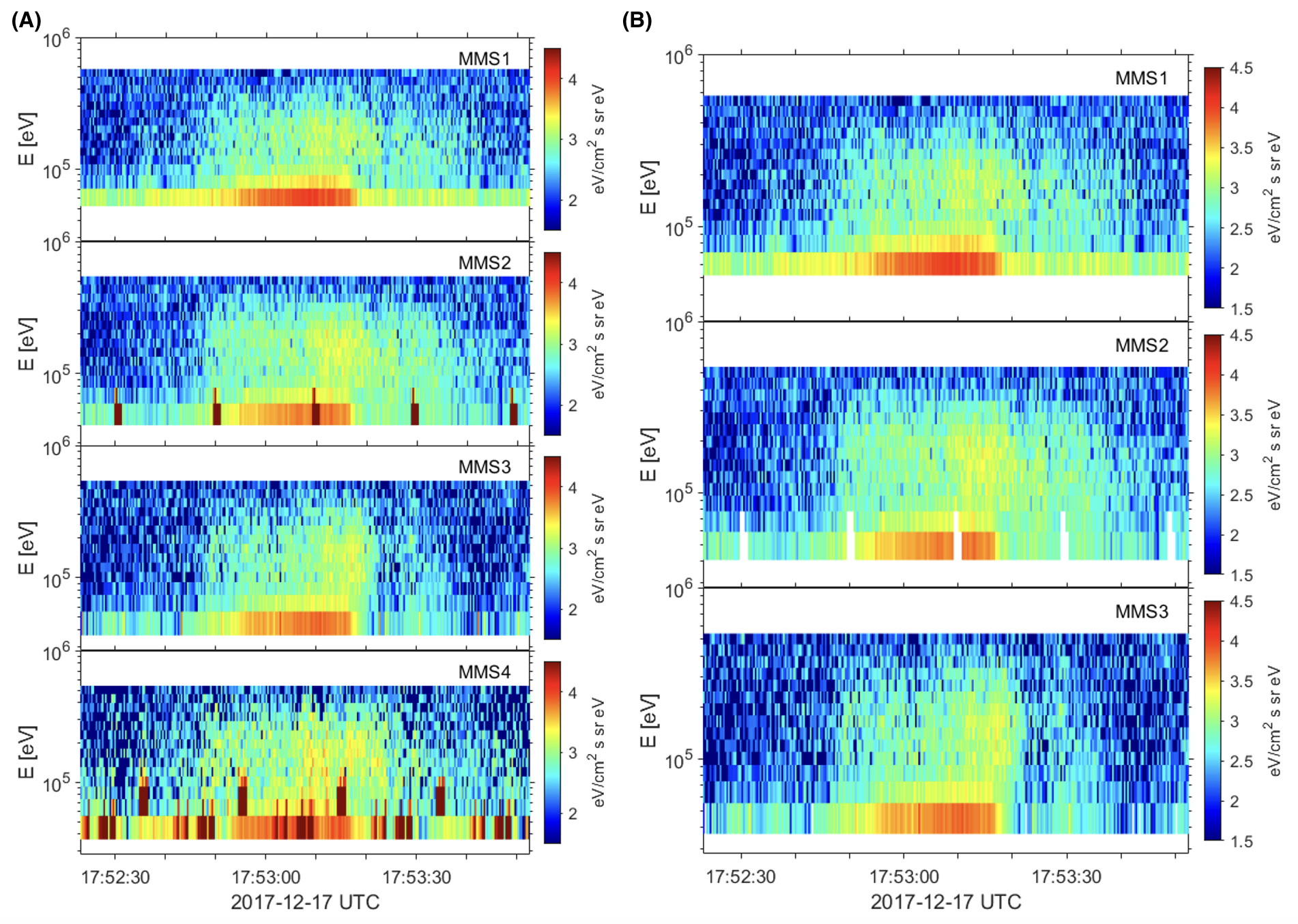}}\caption{
\textbf{Raw FEEPS measurements along with cleaned post-processed products}. The data to the left  \textbf{(A)} show electron differential energy flux from MMS1-4 obtained from the FEEPS instrument in burst resolution. The data to the right \textbf{(B)} are the cleaned data from MMS1-3 which were used in the manuscript. for the generation of Fig. 1.}\label{fig:S1}
\end{figure*}

\noindent\textbf{Data post-process and derivations}. Before using the high-energy electron differential energy flux data, obtained from the FEEPS instrument, the data needs to be cleaned from contamination. This is done by inspecting the measurements from MMS1-4. The MMS4 data typically, and also in our case, has significant contamination which made us choose to discard it in our study (see Fig. \ref{fig:S1}). The contamination can be clearly seen in the last panel of the left column. By looking at the MMS2 data (second panel left), one can see the unphysical high flux values appearing every 20 seconds due to the spin of the spacecraft. The final (cleaned) MMS1-3 data is then shown in the right three panels in Fig. \ref{fig:S1}

The energy bins of FEEPS for each of the MMS spacecraft are slightly different for the electron differential energy flux. Therefore, when combining the cleaned FEEPS electron data of MMS1-3, the energy channels need to be combined properly. Table. \ref{tab:S1} shows the individual spacecraft energy bins and the resulting combined ones.

The noise level of the FEEPS high energy electron spectra (shown in Fig. 1(i, j)) is obtained by a 15-second average window between 17:58:[10-25] on the 2017-12-17 UTC. This set of data is observed to have little to no counts and is located not too far away in time from our event and is therefore ideal to use as a measure of the instrument(s) noise level. The noise level of the FPI electron flux (Fig. 1i) is determined in the same way but for the FPI instrument. However, due to the consistently high-count rate observed at energies below 10 keV, the procedure is only done for energy channels above 10 keV. For energy channels less than 10 keV, the noise level is extrapolated linearly using a linear fit obtained from the observed noise data above 10 keV.

Fig. 1 shows the combined ion spectra of the FPI (6 eV – 30 keV) and FEEPS (60-600 keV) instruments. The gap of no data between 30-60 keV is exponentially interpolated by assuming an exponential decay between the highest FPI channel and the lowest FEEPS channel.

The spectral index calculated in Fig. 1h is calculated by a fitting process on the phase space density (PSD) profiles for energies between 80-200 keV. The PSD profiles are obtained from 2 second averages evenly spaced between 17:52:30 to 17:52:50 using the electron differential energy flux data shown in Fig. 1. The electron energy flux is converted into PSD using the following formula \add{\protect{\cite{hilmer2000enhancement}}}:

\begin{equation}
    \mathrm {\textit{f} = 1.6 \cdot 10^{-40} \cdot \frac{J}{E^2(E+1.02)} \; {[s^3/cm^6]}}
\end{equation}

where E is given in MeV and J in $\mathrm{cm^{-2} s^{-1} sr^{-1}}$.

To determine the electron acceleration efficiency, we need to estimate the total energy carried by solar wind and then which portion is transferred to the suprathermal electrons. The suprathermal electron energy density can be calculated using the expression:

\begin{equation}
    \mathrm{U_e = 4 \pi \sqrt{2/m_e} \int^{E_{max}}_{E_0} \textit{f}_e (E) \sqrt{E^3} dE}
\end{equation}

Where we set the initial suprathermal energy to be $E_0=100$ eV and the maximum to be the maximum energy the FEEPS instrument can measure (515 keV).

For the solar wind it is easy to calculate the energy density upstream of the compressive edge of the foreshock transient structure through the expression $\mathrm{U_i=n_i m_i V_{SW}^2/2}$ , where $\mathrm{n_i}$ is the ion number density, $\mathrm{V_{SW}}$ is the ion speed in the spacecraft reference frame, and $\mathrm{m_i}$ is taken to be equal to the proton mass as it is the dominating population in the solar wind.

To compute these two expressions, we need to take the electron distribution when the spacecraft probes the acceleration region to estimate the suprathermal electron part, and in the upstream region for the solar wind. These regions are shown in detail on Fig. 1. Specifically, they correspond to 2017-12-17 17:53:[15-17] UTC for $\mathrm{U_e}$ and  2017-12-17 17:53:[30-50] UTC for $\mathrm{U_i}$. For the solar wind upstream region, we took the number density to be $\mathrm{n_i=10}$ [$\mathrm{cm^{-3}}$] and $\mathrm{V_{SW}=600}$ km/s which are also typical values of the fast solar wind. By using these values, we obtain the electron acceleration efficiency to be:

\begin{equation}
    \mathrm{v = \frac{U_e}{U_i} \sim 0.05 \sim 5 \%}
\end{equation}

It should be noted that this refers to the typical acceleration efficiency used in the literature. Or in other words the portion of particles that are above the suprathermal range (i.e., 100 eV in our case). If we take $E_0$ to be higher so that it corresponds to relativistic regimes (e.g., 1 keV), the efficiency drops to less than 0.5\%. \\

\noindent\textbf{Solar wind origin determination}. To determine the origin of the solar wind we used the classification scheme described in \cite{xu2015new}. In this work the solar wind is classified in four categories. Ejecta (EJ), Coronal hole (CH), Sector reversal (SR), and Streamer belts (SB).  With CH essentially consisting of fast solar wind plasma, SR and SB indicate slow solar wind conditions, and ejecta are transient phenomena such as Coronal Mass Ejections (CMEs). 

At first, we evaluated the statistical properties of solar wind conditions as measured by both OMNIweb data in 1-h and 1-min resolution. Additionally, to make sure that the classification accurate and does not depend on instrumentation variation, we used \textit{in-situ} observations of ARTEMIS satellite to re-classify the dataset. All three datasets produced identical classification due to the limited variability of the SW between L1 and the lunar orbit.

The actual classification was initially made through visual inspection and by comparing the statistical properties of each class to each event \cite{xu2015new}. However, to make our classification more robust and reproducible, we re-classified the dataset by using a supervised machine learning approach as discussed in \cite{camporeale2017classification}. Specifically, we used the same training set which was carefully curated and analyzed in \cite{camporeale2017classification} and we applied a Gaussian Process (GP) to classify probabilistically the dataset used. All the upstream datasets we used resulted in the same classification (i.e., the highest probability was consistently found to be in the same class). 

\begin{figure*}[ht]
    \centering
{\includegraphics[width=0.7\textwidth]{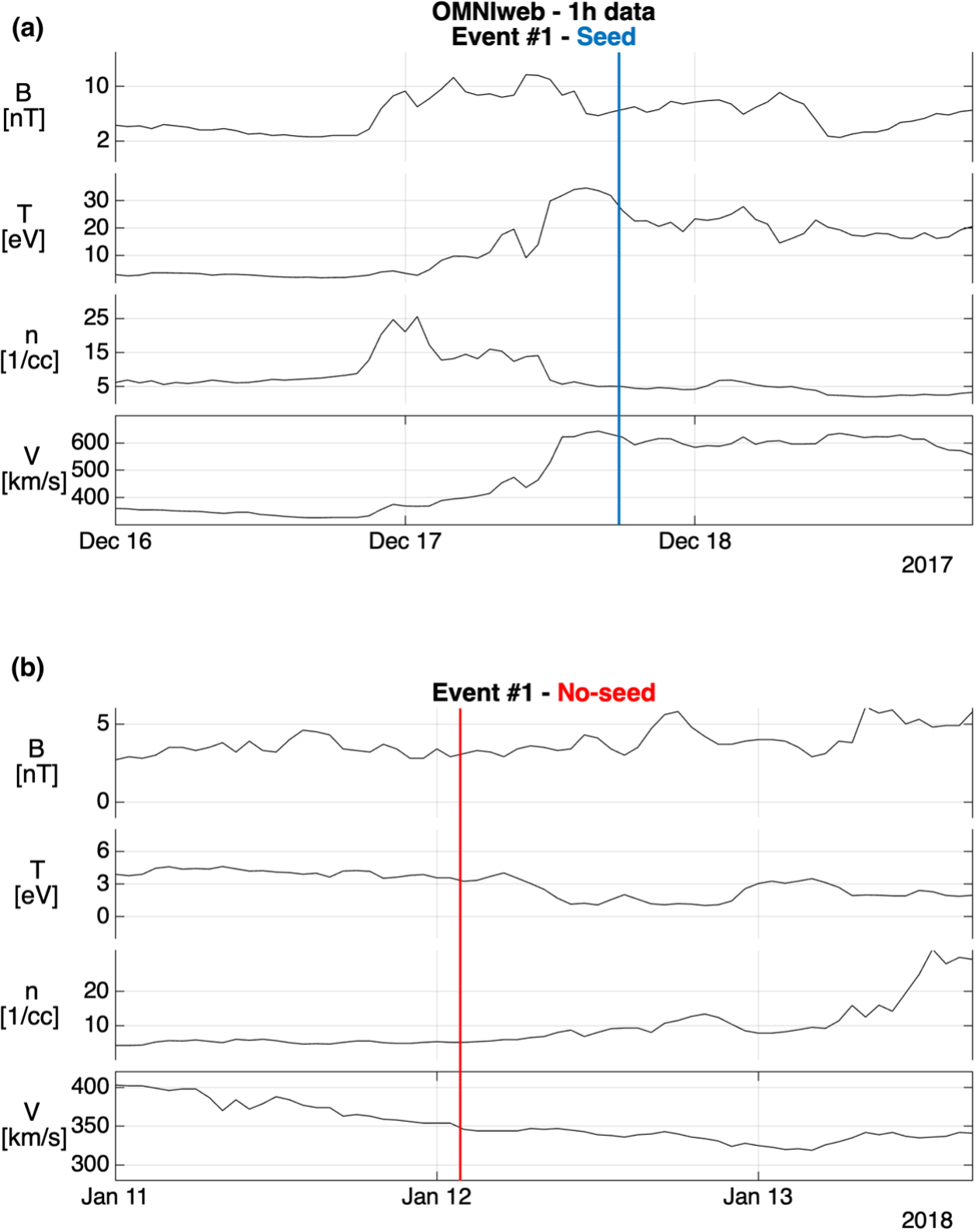}}\caption{
\textbf{Global properties of the solar wind conditions for two events}. (a) shows the conditions for the main seeded event, while (b) shows the conditions for the \#1 non-seeded one (according to the numbering in method section and Fig. 3). From top to bottom, magnetic field absolute value, ion temperature, plasma density, and ion velocity. The vertical lines indicate the time that each foreshock transient was registered at close to earth orbit using MMS. Looking at the global solar wind conditions, we clearly see not only the fast and slow solar wind conditions, but also the fact that the main event is clearly associated to a high-speed stream occurring after the change from slow solar wind ($<$400 km/s) to fast solar wind ($>$400 km/s). On the other hand, the non-seeded event is associated with a totally different solar wind profile of less variable, slow speed conditions.}\label{fig:S2}
\end{figure*}

In Table. \ref{tab:S2}, we show the probabilities of each class for each event, while with bold text one can find the highest probability which was used as the dominant class characterizing each event. Furthermore, in Fig. \ref{fig:S2} OMNIweb 1h data are shown, describing the global picture of the solar wind conditions for the main seeded event and a non-seeded one (i.e., the \#1 non-seeded of Fig. 3). The main event analyzed in the manuscript occurs a few hours after the beginning of a high-speed stream. \\

\begin{figure*}[ht]
    \centering
{\includegraphics[width=0.7\textwidth]{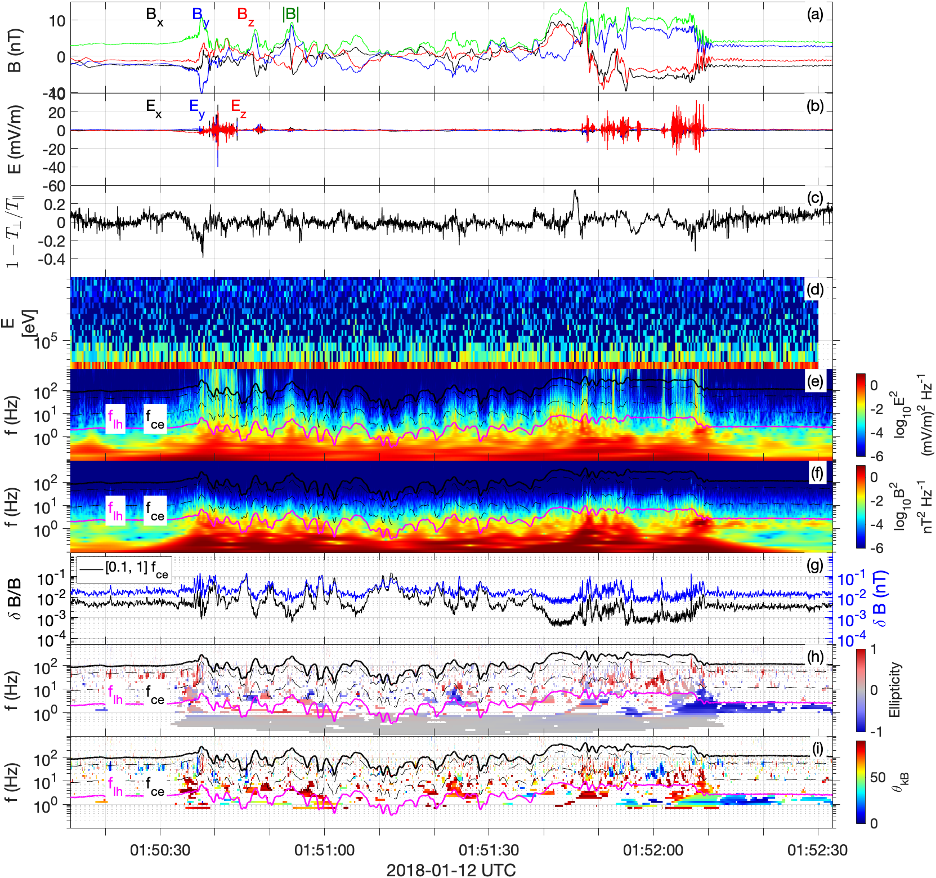}}\caption{
\textbf{Overview plot of the non-seeded event \#1.} (a) magnetic field components and magnitude in GSE coordinates, (b) electric field components and magnitude in GSE coordinates, (c) electron temperature anisotropy, (d) high-energy (FEEPS) differential electron spectra (e) electric field power spectra, (f) magnetic field power spectra, (g) $\delta B/B$ and $\delta B$ timeseries filtered between 0.1 and 1 electron cyclotron frequency (h) ellipticity showing the polarization of the magnetic field power spectra, and (i) wave propagation angle. As shown here, while this foreshock transient is well-formed, the lack of strong electromagnetic activity and seed population resulted in FEEPS measurements (panel d) to essentially measure background noise.}\label{fig:S3}
\end{figure*}

\noindent\textbf{Diffusion time scales and spatial scale sizes}. To estimate the diffusion time scales and compare them with the system scale sizes, we have used Bohm’s diffusion formula \cite{caprioli2014simulations}. This formula is applicable when we have variations in the magnetic field that are in the same order as the background magnetic field values (i.e., $\delta \mathrm{B} \sim \mathrm{B_0}$). As shown in Fig. 1 and 2 and Fig. \ref{fig:S3}, this is a systematic feature of the foreshock transients’ core environment (i.e., the region between the trailing and leading compressive edge). Under this framework, the mean free path for pitch angle scattered electrons can be computed as $\lambda_{max} \sim r_{ce}$ where $r_{ce}$ is the electron gyroradius. 

The diffusion length (L) can be estimated through the expression $\mathrm{L \sim  D(E_{max})/V_{sh}}$ , where $\mathrm{D(E_{max})}$ is the spatial diffusion coefficient for the maximum energy, and $V_{sh}$ is the velocity of the shock in the spacecraft reference frame.  Finally, the diffusion coefficient can be calculated through $\mathrm{D(E_{max} )=1/3 \cdot v \cdot \lambda_{max}}$, where v is the velocity of the particles. Combining the above formulas, one can easily obtain an expression for the maximum diffusion length as:

\begin{equation}
    \mathrm{L \sim 1/3 \cdot v/V_{sh} \cdot r_{ce}}
\end{equation}

In the above formula, we can use the relativistic kinetic energy expression, $E=(\gamma-1)\mathrm{mc^2}$, where m is the rest mass, c is the speed of light, and $\gamma$ is the Lorentz factor. By expanding the Lorentz factor, and re-writing the kinetic energy expression, one can show that:
\begin{equation}
    \mathrm{v = c \sqrt{1 - \left(\frac{mc^2}{E_{max}+mc^2} \right)}}
\end{equation}

Using that expression, since we know the maximum energy, we can calculate the equivalent velocity and translate that to an effective electron gyroradius though $\mathrm{r_{ce}=\frac{\gamma m v}{eB_0}}$ where e is the electric charge, and $\mathrm{B_0}$ is the background magnetic field. Therefore, observationally, we can obtain a value for the electron gyroradius by measuring the energy of electrons and the background magnetic field. Finally, to compute the diffusion length, one needs to estimate the shock velocity. One can do that by timing the shock as it passes through the multiple MMS spacecraft. By performing the timing method \cite{turner2021direct,amano2020observational}, we obtain that the shock velocity in the spacecraft reference frame to be in the order of  $\mathrm{V_{sh} \sim300}$ km/s. 

The background magnetic field at the core of a foreshock transient is in the order of 1 nT as shown in Fig 1, 2 and \change{S4}{7}, although in extreme cases it can reach values up to 10 nT. The maximum energy of electrons observed in our event is 523 keV for the main event analyzed in the manuscript. By using the above values, we obtain a diffusion length of  $\mathrm{L \sim 10-100 \; R_e  \sim 5\cdot 10^{4-5}}$  km, which is in agreement with the scale sizes foreshock transients \add{that }have been \add{observed }at Earth’s planetary environment (with 100 $R_e$ corresponding to extreme cases) \cite{valek2017hot}.

In order now to \remove{roughly} estimate what \add{are} the maximum \add{obtainable} energies in other planetary systems, we can take the typical scale sizes found in other planets along with typical background magnetic field observations \cite{valek2017hot}. This energy basically corresponds to the maximum energy that an electron can have before their gyroradius gets so large that they cannot remain in the shock acceleration region and get further energized. We can assume that for giant planets like Jupiter, a foreshock transient scale size can reach values of  $\mathrm{\sim 100-1000 \; R_e  \sim 5 \cdot10^{5-6}}$  km. Using a background magnetic field value of 0.5-1 nT, it can be easily shown that the equivalent energy obtained from the electrons' gyroradius can be $>2$ MeV.

Finally, to generalize our result to systems outside our solar system, some assumptions need to be made. Generally, the larger the system size (planet size) the larger the maximum energy that can be sustained. Similarly, the stronger the magnetic field of the stellar wind, the higher the energy. As a result, one could investigate ultra-hot Jupiters as potential candidates in which electrons can achieve extremely high energies. These celestial objects are Jupiter like exoplanets with very low orbital periods due to their proximity to their star \cite{cauley2019magnetic}. 

If we assume a system size similar to Jupiter, we can obtain a foreshock transient scale size can reach values of $\mathrm{\sim100-1000  \;R_e  \sim 5 \cdot10^{5-6}}$  km. Now depending on the age of the stellar system (young stellar systems have stronger magnetic field embedded in their stellar wind), one can estimate the background magnetic field magnitude of the stellar wind \change{in close to}{near a planet close to} the star \remove{orbits} to be in the order of $\mathrm{B \sim 0.01-1 \; G \sim 10^{3-5}}$ nT \cite{nichols2016stellar,chebly2023numerical}. By following the exact argumentation as above, we can obtain that the maximum energy in these systems can be in the order of GeV to TeV.  It should be noted that even when computing the synchrotron losses for these energy ranges and magnetic fields, the loss is insignificant. As observations and modeling efforts in these systems are highly susceptible to errors, one should treat the whole discussion here not as a direct comparison to other acceleration mechanisms but rather as a \change{direct result}{consequence} of our model that needs further investigation by the astrophysics and exoplanet research communities. \\

\noindent\textbf{Foreshock transient \& shock characterization}. 

While the characterization of the foreshock transients is out of the scope of this work, it should be noted that the majority of the transients shown in this work would have been classified as hot flow anomalies (HFAs) in heliophysics terminology. The exact characterization of these transients is out of the scope of this work and irrelevant to our findings as regardless of the exact nature, foreshock bubbles and HFAs have been found to be extremely efficient accelerators (see e.g., \cite{wilson2016relativistic,liu2017fermi,liu2019relativistic}). Furthermore, both types of phenomena  contain all the necessary ingredients for our model (i.e., shock formation, compressive edges, high-frequency nonlinear waves, large amplitude waves associated to low density region to facilitate scattering etc.) \cite{zhang2022dayside}.

To characterize the shock formed at the compressive edge of the foreshock transient (highlighted in Fig. 1 and 2), we used single-point and multipoint analysis techniques (see e.g., \cite{turner2021direct}). All techniques as expected from previous results indicated that the fast shock formed at the sun-facing compressive edge of the foreshock structure is a typical quasi-perpendicular ($\mathrm{\theta_{Bn}>45^\circ}$) shock. Specifically, all the methods used (i.e., coplanarity theorem, timing, and minimum variance analysis) showed an angle systematically above 55 degrees similar to other similar foreshock transient events \cite{turner2021direct}. \\

\noindent\textbf{Wavelet and wave-particle interaction analysis}. For Fig. 2, we calculate the power spectral density (PSD) of the electric and magnetic fields using a wavelet transform, which allow us to observe how the power of the various oscillations is distributed in frequency and how it varies with time. Furthermore, to characterize the various wave modes excited we use singular value decomposition (SVD) to calculate the degree of polarization, ellipticity, and $\mathrm{\theta_{kB}}$, which are respectively a measure of the coherence of the waves, a measure of the handedness of the waves, and the angle between the wave vector (k) and the background magnetic field ($\mathrm{B_0}$). The ellipticity has a value in the interval [-1,1] where values of -1, 0, and 1 reflect waves that are left hand, linearly, or right hand polarized respectively. The propagation angle ($\mathrm{\theta_{kB}}$) ranges between 0 and 90 degrees with 0/90 representing parallel/ perpendicular propagations with respect to $\mathrm{B_0}$.

The identification of the high-frequency whistler waves was mainly obtained by their typical frequency that lies within [0.1 1]$f_\mathrm{{ce}}$ with $f_\mathrm{{ce}}$ being the electron cyclotron frequency. To quantify the relative power of the wave activity, we integrated the PSD in frequency space for each interval of interest (compressive region of the foreshock transient). This approach allowed us to estimate the total magnetic field variability for the specific wave-mode ($\mathrm{\delta B^2}$). To evaluate how prominent the wave amplitude is compared to the background magnetic field and essentially quantify the nonlinearity of the wave, we took the square root and divided by the background field to obtain an estimate of $\mathrm{\delta B/B}$ which was used in Fig. 3. The exact values obtained from this methodology are shown in Table. \ref{tab:S3}.

To evaluate the cyclotron resonance between the energetic electrons and the whistler waves, we need first to write the resonance condition: $\mathrm{\omega-k_z v_z+m \Omega_{ce} = 0}$ \cite{breech2009electron}, where $\mathrm{\omega}$ is the frequency of the wave, while $\mathrm{k_z}$ and $\mathrm{v_z}$ are the components of the wave vector and the electron velocity along the background magnetic field respectively. Finally, $\mathrm{\Omega_{ce}}$ is the electron gyrofrequency. In the main event presented, we estimate $\mathrm{\omega \sim 0.3 \Omega_{ce}}$ which results in a frequency $f_{ce} \sim 500$ Hz \add{calculated via $ f= \frac{\omega}{2\pi}$}. By using the cold plasma dispersion relations, the observed high-frequency whistlers have a wavelength of $\lambda \sim 25$ km. For the observed electrons a realistic initial energy of 20 keV, the total velocity of the particles is estimated to be $\mathrm{v_e \sim 8.4 \cdot 10^7}$  km/s. For such electrons to be in cyclotron resonance with the observed HF whistlers, the pitch angle (PA) of these electrons needs to be $\sim 79^\circ$. Similarly, for a slightly higher energy of 50 keV, their PA needs to be $\sim 83^\circ$. The interaction between the electrons is shown in Fig. 3., as the nonlinear resonance of the electrons will make them gain energy in the perpendicular direction which will further increase their PAs reaching asymptotically $90^\circ$ \cite{thorne2013rapid,horne2003resonant}. As discussed in the Results section, during the strongest whistler activity, we see in the Pitch Angle distribution (Fig. 2. panel h) an enhanced flux at PA $90^\circ$ which is a \remove{clear} signature of the above-mentioned resonant energization mechanism. \\

We should also note the acceleration process via wave-particle interaction discussed in this work is not related to quasi-linear theory and pitch angle diffusion (e.g., \cite{allison2021gyroresonant}). \change{and as written in the main manuscript it discusses}{As written in the main manuscript, the resonance we highlight is between} nonlinear \add{waves and electrons } \remove{interaction} (e.g., \cite{bortnik2008nonlinear}).  \add{The nonlinear nature of these waves is evident from their relatively high amplitude compared to the background (see Fig. 3 and Table III) and their characteristic coherent narrowband emission \protect\cite{zhang1999whistler,artemyev2022electron}}. Having said that, acceleration mechanisms through resonance are found to be more efficient for low density plasma with high magnetic field (e.g., radiation belts). It has been shown that a way to quantify the efficiency of the acceleration obtained through this mechanism is to look at the ratio between the electron plasma frequency ($f_\mathrm{{pe}}$) and gyrofrequency ($f_\mathrm{{ce}}$). The ratio in radiation belt can be close to unity (see e.g., \cite{allison2021gyroresonant}) while in typical foreshock plasma it can be order of magnitudes higher. However, due to the special plasma conditions of foreshock transients, the core (Fig. 1, blue shaded region) contains very low density plasma, while the shock (Fig. 1, green shaded area) contains magnetic field that due to compression is significantly higher than the typical foreshock. These combined allow this ratio for our main event to be in the order of $f_\mathrm{{pe}}/f_\mathrm{{ce}} \sim30$ when considering the bulk properties. Considering partial distributions which are more accurate in shock/foreshock environments should decrease the density and therefore also the fpe/fce ratio (see e.g., \cite{raptis2022downstream}) even more. \add{Furthermore, although the efficiency of wave-particle interactions may not be as pronounced as in very low-density plasma environments, recent studies have demonstrated that these interactions can still significantly contribute to the energization and scattering of electrons in a foreshock transient \protect\cite{shi2023intense,vargas2023role}.}Moreover, when moving to exoplanetary systems with stronger magnetic field (e.g., ultra-hot Jupiters \cite{cauley2019magnetic}) this ratio is significantly decreased which further supports the generalization of our model to other environments. \add{Finally, while it remains uncertain whether interactions between electrons and high-frequency whistlers will average out to a quasi-linear picture, the nonlinear characteristics observed in the waves suggest deviations from classical resonance theory, given that for all events we observe a $\delta B/B \sim 1-10\%$}.

\noindent\textbf{Occurrence and probability analysis of foreshock transients}. As briefly mentioned in the main text, foreshock transient phenomena can be present in every planetary environment within a stellar system, and occur very frequently \cite{schwartz2000conditions}. However, obtaining actual \textit{in-situ} observations in our solar system of all the elements of the presented mechanism is heavily influenced by various statistical and orbital effects, qualitatively discussed below.

The first statistical issue affecting our observational capabilities is the probability to observe fast solar wind (coronal hole) plasma corresponding to the conditions we verified as necessary to have a prominent electron seed population (1-5 keV range), highlighted throughout the main text. Coronal hole plasma consists of roughly 20\% of the total solar wind measurements based on Lagrangian 1(L1) point observations (\cite{xu2015new,borovsky2016plasma,camporeale2017classification}). Additionally, fast solar wind is associated with more frequent magnetic field discontinuities \cite{tsurutani1999review}, and the occurrence rate of such a discontinuity is roughly in the order of 1-10 every hour \cite{liu2022magnetic}. These effects show that while foreshock transients can form with an occurrence of hours \cite{schwartz2000conditions}, seeded events exhibiting significant electron acceleration like the one we describe may have an occurrence rate in the order of days. 

Having said that, from an observational point of view, to measure the seeded population (ARTEMIS) and the foreshock transient (MMS), both spacecraft need to be at a specific location of their orbit.

MMS spacecraft is in the ion foreshock for around 10\% of its lifetime, although this heavily varies based on its orbit. Furthermore, in order to do the analysis required in this manuscript, MMS has to acquire high time resolution measurements (i.e., burst data) that are typically unavailable during foreshock observations, making this percentage significantly lower. Moreover, to observe these phenomena, it is not just sufficient to be in the foreshock at the right time, but also to observe the foreshock transient after it has been fully formed, which reduces the occurrence even more. Moving on to ARTEMIS mission in lunar orbit, it can provide far upstream measurements for less than 50\% of the time, as for the rest of the time it is on the nightside of the Earth or very close to the shock. This results in a net probability of observing such a foreshock transient to be less than 1-5\% even considering the most conservative estimations. As a result, while these phenomena can occur very frequently \cite{schwartz2000conditions}, the observational capabilities we currently have, allow us to observe roughly a few of such conjunction events per year.

All the effects listed above are the reason that a detailed investigation of several years of data resulted in 6 seeded cases that were analyzed in Fig. 3. \add{Our research targets the detection of high-energy electrons, enabling us to identify specific acceleration events. However, it is crucial to recognize that non-energetic events can still occur under the influence of coronal hole (CH) solar wind. Factors such as suboptimal geometry, limited wave-particle interactions, and insufficient compression can constrain substantial electron acceleration. As a result, significant electron acceleration might not always be observed, even in the presence of CH solar wind.}

\noindent\textbf{Statistical analysis details}. In this work we focused on analyzing the most energetic event (i.e., the one shown in Fig. 1 and 2 and \#1 seeded in Fig. 3) which highlights all the components of the presented acceleration model. However, for the statistical verification of our work, we examined multiple cases that contained partially or fully the components of our model (Fig. 3). On Fig. \ref{fig:S3}, we show an overview plot of an event that was compared with the main event (i.e., \#1 non-seeded on Fig. 3). There, one can see that a well-formed foreshock transient that lacks the initial seeded population from the solar wind results in no observable  electron acceleration (Fig. \ref{fig:S3} panel d). This is also the event that was compared side by side with the main event on Fig 4.

 Furthermore, for verification and to make our work reproducible, a table with all the dates for each event as measured by MMS along with the associated ARTEMIS upstream observations are provided (Table. \ref{tab:S4}). The intervals shown there were used to obtain the maximum flux ratio compared to the surrounding that was used in Fig. 3 and 4.

It should be noted that for the determination of the magnetic field discontinuity responsible for the formation of each foreshock transient, we used a multi-step process. First, based on the position of the two spacecraft, we estimated an approximate time lag by using the solar wind velocity as measured by MMS. Then we determined whether a magnetic field rotation was observed before and after the foreshock transient at MMS. Using the rotation associated to the foreshock transient, we proceed to cross-validate our findings with the magnetic field rotations occurring within the time lagged interval at ARTEMIS dataset. This resulted in an associated discontinuity at ARTEMIS orbit occurring from 2 to up to 20 minutes before the MMS observations. After confirming the magnetic disturbance per event, we limited the time interval to be within $\sim$5 minutes of the observations and obtained the background value by evaluating steady solar wind calm conditions within a 20-minute window. As shown in Table \ref{tab:S4}, the relative position of the MMS with respect to ARTEMIS can have a wide range of values. This along with with the fact that the orientation of the normal of the discontinuity can be at the same plane as the separation vector between MMS and ARTEMIS, allows the time lag in certain cases to be relatively small, or even instantaneous (e.g., event \#5). Since the discontinuity can interact with Earth's bow shock and form a foreshock transient at the same time as observed by ARTEMIS, these cases are statistically expected. 

At this point we should stress that for the events occurring on 2018-12-10 (i.e., \#3, \#5, \#6) the high-energy electron instrument of ARTEMIS (SST) observed $\sim 100$ keV electrons. For the rest of the events, either the data quality of the instrument showed faulty data and/or no high-energy electrons were measured. This suggests that the electrons accelerated from the presented mechanism, as expected, can be scattered and  travel both sunward and earthward.

Finally, as mentioned in the main text, apart from the solar origin, the seed population can be associated with the presence of an extended electron foreshock. An exclusive foreshock origin seems unlikely since the electron foreshock is always present, and this could not explain why in well developed transients (e.g., Fig. \ref{fig:S3}), with clear shock formation and wave activity there is no significant particle acceleration. However, it is likely that electron foreshock can contribute to the seed population as reflected particles from the planetary bow shock can get \change{scaterred}{scattered} and interact with the upstream foreshock transient (see e.g., relevant articles on ions \cite{liu2017fermi,turner2018autogenous}).

\begin{figure*}[ht]
    \centering
{\includegraphics[width=1.0\textwidth]{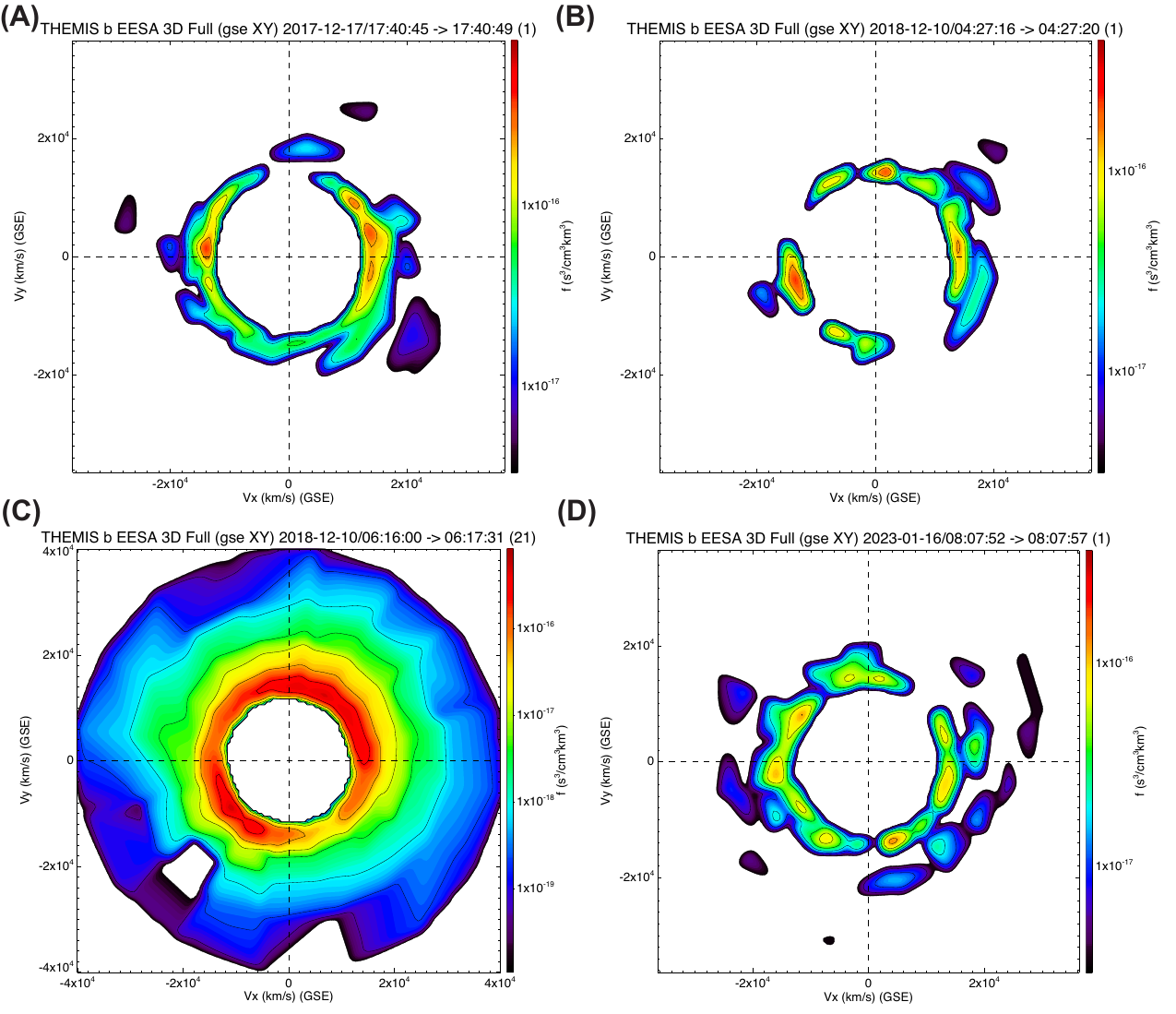}}\caption{
\textbf{2D distribution slices for selected ARTEMIS events}. 2D distribution slices in velocity space in GSE coordinates, showing ARTEMIS observations for three seeded and one non-seeded events (namely, (A)  for \#1,  (B) for \#6, (C) for \#3 and (D) for \#2 no-seed).  the X axis shows the velocity along X GSE coordinates (Earth-Sun line). These ring like distributions indicate that the electrons population is travelling towards and from the Earth. The slices are obtained by restricting the energy between 500 and 5000 eV in order to be directly associated to the seed population discussed in the manuscript.}\label{fig:S4}
\end{figure*}

To evaluate this possibility we first investigated the partial distribution function for the \add{low keV} seed population observed at ARTEMIS orbit \add{(Fig. 8)}. For all events we see particles between the seed energy ($\sim 1-5$ keV) to form a ring-like distribution in the XY GSE plane. This type of distribution could be considered a part of a Bi-Maxwellian like distribution that hint towards a clear solar origin. However, it could be part of the electron distribution originating from particles reflected from the nearby Earth's bow shock (i.e., electron foreshock). This, while remains a possibility, is not very likely. By examining the distribution one can see that the events exhibiting these features are associated with low statistics both due to the instrument mode having a low sampling frequency but also due to the limitation on energy range \cite{angelopoulos2014artemis}. This is illustrated in Fig. \ref{fig:S4} (C) where an event with another instrument mode activated allows more frequent measurements indicates the presence of a typical thermalized, SW type distribution being present. Another test is to compare the distribution found in seeded events with the non-seeded ones (e.g., Fig. \ref{fig:S4} (D)). As shown there, the distributions of the seeded events (panels A-B) are very similar to the ones with the non-seeded with the exception of enhanced PSD which appears to occur for both earthward and sunward particles. This hints towards a complex interplay in which both solar and foreshock origin particles may be present. \add{It is important to note that beam-like distributions are evident in the seeded events, as depicted in Figure 8 (panels A-B). This observation indicates that, in addition to potential contributions from the foreshock, instabilities within the solar wind may also contribute to the formation of the seed population \protect\cite{lin1998wind,verscharen2022electron}}.

\begin{figure*}[ht]
    \centering
{\includegraphics[width=1.0\textwidth]{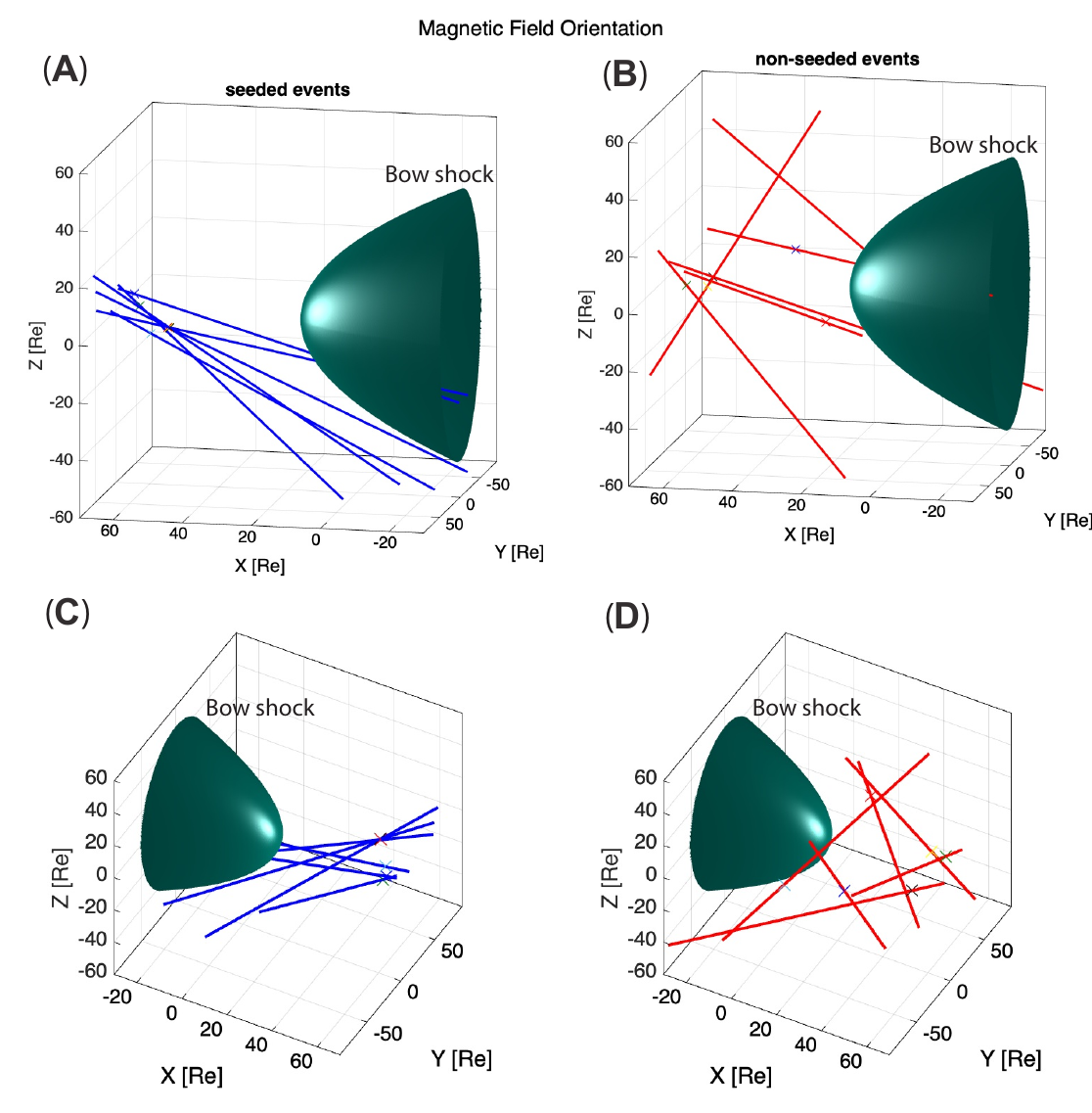}}\caption{
\textbf{Bow shock modeling and magnetic connectivity}. 3D bow shock model using \cite{chapman2003three} under typical solar wind conditions. (\textbf{A}) shows the location of ARTEMIS and the magnetic field lines in GSE coordinates during the maximum electron flux for the suprathermal-seed energy range (see Table \ref{tab:S4}). As shown here, only 2 out of the 6 seeded events appear to be connected to the bow shock.  (\textbf{B)} the same plot but for the non-seeded event.  (\textbf{C-D)} follow the same format but from a different point of view. Here one out of 6 magnetic field lines is connected to the Earth's bow shock.}\label{fig:S5}
\end{figure*}

\change{In order to determine how likely this we can continue this investigation by evaluating the magnetic connectivity of all events between ARTEMIS and the Earth's bow shock}{Our next step is to investigate the linkage between solar wind observations by ARTEMIS and Earth's bow shock. We aim to determine whether a systematic connection to the electron foreshock exists across all seeded events. If such a consistent connection is found, it would imply that the reflected foreshock population may be the primary mechanism driving the observed seed population}. \change{By using}{Through} a 3D bow shock model (\cite{chapman2003three}) we used the magnetic field orientation for both seeded and non-seeded events and investigated whether they intersect the bow shock surface.  As shown in Fig. \ref{fig:S5}, 2 out of 6 seeded events and 1 out of 6 non-seeded events appear to be unambiguously mapped to the bow shock. For the seeded events we took the magnetic field orientation during the peak of the suprathermal electron flux. For the non-seeded events, since there is no significant enhancement in electron flux, we took the magnetic field orientation before the discontinuity (taking the magnetic field data after the discontinuity produced the same results). The analysis shown on Fig. \ref{fig:S5} shows that electron foreshock cannot be the only source of suprathermal electrons. It does show however, that seeded events that experience significant acceleration of electrons have discontinuities for which the geometry is optimal for the formation of foreshock transients. The orientation of these field lines (Fig. \ref{fig:S5} (A)) essentially allows the intersection point with the shock to slowly move across the bow shock as the field lines get convected with the nominal solar wind. This illustrates the geometrical effect that the foreshock transients can have further enhancing the acceleration process discussed in the main text and illustrated in Fig. \ref{fig:4}.

\noindent\textbf{Data availability}. The MMS data are archived at \url{https://lasp.colorado.edu/mms/sdc/public/}. The THEMIS/ARTEMIS data can be found at \url{https://themis.ssl.berkeley.edu/data\_products/index.php}, while the OMNIweb data is accessed through \url{https://spdf.gsfc.nasa.gov/pub/data/omni/}.\\

\noindent\textbf{Code availability}. The analysis of the work was done via the PySPEDAS (\url{https://github.com/spedas/pyspedas/tree/master}) \cite{grimes2022space}, SPEDAS (\url{http://spedas.org/blog/}\cite{angelopoulos2019space} and IRFU-Matlab (\url{https://github.com/irfu/irfu-matlab/tree/master}) \cite{khotyaintsev_2024_11550091} libraries. Specifically PySPEDAS was used to download the observations and IRFU-Matlab to analyze and process the files for the plots and the analysis shown in the manuscript. One can access the version of the codes used along with instructions on how to reproduce each figure and table of our work on the associated Zenodo repository \cite{savvas_2024_13619432}.\\

\clearpage
\noindent\emph{References}

\clearpage
\noindent\emph{Acknowledgments}

 This work was supported by the Magnetospheric Multiscale (MMS) mission of NASA’s Science Directorate Heliophysics Division via subcontract to the Southwest Research Institute (NNG04EB99C). SR acknowledges funding from NASA DRIVE Science Center for Geospace Storms (CGS) — 80NSSC22M0163, and Johns Hopkins University Applied Physics Laboratory independent R\&D fund. SR acknowledges the support of the International Space Sciences Institute (ISSI) team 555, “Impact of Upstream Mesoscale Transients on the Near-Earth Environment”, and the Archival Research Visitor Programme of the European Space Agency (ESA).  AL acknowledges funding from Swedish Research Council grant 2018-05514. ML acknowledges the support of the International Space Science Institute (ISSI) in Bern, through ISSI International Team Project 520 and funding from Swedish Research Council grant 2018-05514. DLT is thankful for funding from NASA’s MMS and IMAP missions. DLT also acknowledges funding from National Science Foundation (NSF) Geospace Environment Modeling (GEM) program 2225463. DC acknowledges funding from NASA H-TMS grant 80NSSC20K1273, NASA H-TMS grant 80NSSC24K0173 and NSF--DoE grant PHY-2010240. We acknowledge the useful discussions with Anton Artemyev, and Florian Koller. We also acknowledge Vassilis Angelopoulos, the instrument PIs and the whole team for use of data from the THEMIS/ARTEMIS Mission. Finally, We acknowledge the entire Magnetospheric Multiscale team and instrument PIs for data access and support.\\

\noindent\emph{Author contributions}\\
Conceptualization:  SR, AL, ML, DLT \\
Methodology: SR, AL, ML, DLT, DC\\
Investigation: SR, AL, ML, DLT, DC\\
Visualization: SR, AL, ML, DLT\\
Funding acquisition: SR, AL, ML, DLT, DC, JLB\\
Writing – original draft: SR \\
Writing – review \& editing: SR, AL, ML, DLT, DC, JLB\\

\noindent\emph{Competing interests}
The authors declare no competing interests.

\renewcommand{\tablename}{Table.}
\renewcommand{\figurename}{Fig.}

\begin{table*}
 \caption{ The different energy channels measuring electron energies for MMS1, MMS2 and MMS3. The two rightmost columns show the combined energy channels (MMS1-3) and which individual MMS channels have been combined to form a common energy bin. The dash (-) sign indicate that no channel (or value) was used \label{tab:S1}}
\begin{tabular}{|l|c|c|c|c|c|}
\hline
\multicolumn{1}{|c|}{\textbf{Channel}} & \textbf{\begin{tabular}[c]{@{}c@{}}MMS1\\ (keV)\end{tabular}} & \textbf{\begin{tabular}[c]{@{}c@{}}MMS2 \\ (keV)\end{tabular}} & \textbf{\begin{tabular}[c]{@{}c@{}}MMS3 \\ (keV)\end{tabular}} & \textbf{\begin{tabular}[c]{@{}c@{}}MMS1-3 \\ (keV)\end{tabular}} & \textbf{Combined Channels} \\ \hline
1                                      & 61.2                                                          & 31.2                                                           & 27.2                                                           & -                                                                & (-,-,-)            \\ \hline
2                                      & 79.9                                                          & 49.9                                                           & 45.9                                                           & 47.9                                                             & (-,2,2)            \\ \hline
3                                      & 98.6                                                          & 68.6                                                           & 64.6                                                           & 64.8                                                             & (1,3,3)            \\ \hline
4                                      & 117.4                                                         & 87.4                                                           & 83.4                                                           & 83.6                                                             & (2,4,4)            \\ \hline
5                                      & 135.1                                                         & 105.1                                                          & 101.1                                                          & 101.6                                                            & (3,5,5)            \\ \hline
6                                      & 153.2                                                         & 123.2                                                          & 119.2                                                          & 119.9                                                            & (4,6,6)            \\ \hline
7                                      & 174.5                                                         & 144.5                                                          & 140.5                                                          & 140                                                              & (5,7,7)            \\ \hline
8                                      & 199.3                                                         & 169.3                                                          & 165.3                                                          & 169.7                                                            & (7,8,8)            \\ \hline
9                                      & 228.2                                                         & 198.2                                                          & 194.2                                                          & 197.2                                                            & (8,9,9)            \\ \hline
10                                     & 262                                                           & 232                                                            & 228                                                            & 229.4                                                            & (9,10,10)          \\ \hline
11                                     & 301.4                                                         & 271.4                                                          & 267.4                                                          & 266.9                                                            & (10,11,11)         \\ \hline
12                                     & 347.4                                                         & 317.4                                                          & 313.4                                                          & 315.4                                                            & (-,12,12)          \\ \hline
13                                     & 401.2                                                         & 371.2                                                          & 367.2                                                          & 369.2                                                            & (-,13,13)          \\ \hline
14                                     & 464                                                           & 434                                                            & 430                                                            & 432                                                              & (-,14,14)          \\ \hline
15                                     & 537.2                                                         & 507.2                                                          & 503.2                                                          & 515.9                                                            & (15,15,15)         \\ \hline
\end{tabular}
\end{table*}

\begin{table*}[]
\caption{Probabilities for each classification made for the presented events (Fig. 3) through a Gaussian process (GP) supervised machine learning model (30). Each cell indicates the probability in decimal number format, while with bold text, the most dominant class is highlighted. For this table, we used the OMNI 1-h dataset. \label{tab:S2}}
\begin{tabular}{|l|c|c|c|c|}
\hline
{\color[HTML]{1F497D} \textbf{}}   & \textbf{\begin{tabular}[c]{@{}c@{}}Ejecta \\ (EJ)\end{tabular}} & \textbf{\begin{tabular}[c]{@{}c@{}}Coronal Hole \\ (CH)\end{tabular}} & \textbf{\begin{tabular}[c]{@{}c@{}}Sector Reversal \\ (SR)\end{tabular}} & \textbf{\begin{tabular}[c]{@{}c@{}}Streamer Belts \\ (SB)\end{tabular}} \\ \hline
{\color[HTML]{1F497D} \#1 seed}    & 0.07                                                            & \textbf{0.78}                                                         & 0.09                                                                     & 0.06                                                                    \\ \hline
{\color[HTML]{1F497D} \#2 seed}    & 0.10                                                            & \textbf{0.40}                                                         & 0.15                                                                     & 0.23                                                                    \\ \hline
{\color[HTML]{1F497D} \#3 seed}    & 0.07                                                            & \textbf{0.79}                                                         & 0.09                                                                     & 0.05                                                                    \\ \hline
{\color[HTML]{1F497D} \#4 seed}    & 0.13                                                            & \textbf{0.39}                                                         & 0.12                                                                     & 0.37                                                                    \\ \hline
{\color[HTML]{1F497D} \#5 seed}    & 0.07                                                            & \textbf{0.77}                                                         & 0.10                                                                     & 0.07                                                                    \\ \hline
{\color[HTML]{1F497D} \#6 seed}    & 0.06                                                            & \textbf{0.79}                                                         & 0.09                                                                     & 0.06                                                                    \\ \hline
{\color[HTML]{C00000} \#1 no-seed} & 0.02                                                            & 0.09                                                                  & 0.09                                                                     & \textbf{0.80}                                                           \\ \hline
{\color[HTML]{C00000} \#2 no-seed} & 0.35                                                            & 0.18                                                                  & 0.05                                                                     & \textbf{0.42}                                                           \\ \hline
{\color[HTML]{C00000} \#3 no-seed} & \textbf{0.56}                                                   & 0.06                                                                  & 0.14                                                                     & 0.23                                                                    \\ \hline
{\color[HTML]{C00000} \#4 no-seed} & 0.13                                                            & 0.12                                                                  & 0.31                                                                     & \textbf{0.44}                                                           \\ \hline
{\color[HTML]{C00000} \#5 no-seed} & 0.20                                                            & 0.01                                                                  & \textbf{0.63}                                                            & 0.16                                                                    \\ \hline
{\color[HTML]{C00000} \#6 no-seed} & 0.25                                                            & 0.05                                                                  & 0.24                                                                     & \textbf{0.46}                                                           \\ \hline
\end{tabular}
\end{table*}

\begin{table*}[]
\caption{\add{Wave amplitude with respect to the background magnetic field for each of the event shown in Fig. 3.} \label{tab:S3}}
\begin{tabular}{|l|c|}
\hline
{\color[HTML]{1F497D} \textbf{}}   & \textbf{$\delta B/B$} \\ \hline
{\color[HTML]{1F497D} \#1 seed}    & 0.0596      \\ \hline
{\color[HTML]{1F497D} \#2 seed}    & 0.0374      \\ \hline
{\color[HTML]{1F497D} \#3 seed}    & 0.0420      \\ \hline
{\color[HTML]{1F497D} \#4 seed}    & 0.0570      \\ \hline
{\color[HTML]{1F497D} \#5 seed}    & 0.0779      \\ \hline
{\color[HTML]{1F497D} \#6 seed}    & 0.1314      \\ \hline
{\color[HTML]{C00000} \#1 no-seed} & 0.0145      \\ \hline
{\color[HTML]{C00000} \#2 no-seed} & 0.0789      \\ \hline
{\color[HTML]{C00000} \#3 no-seed} & 0.0209      \\ \hline
{\color[HTML]{C00000} \#4 no-seed} & 0.0597      \\ \hline
{\color[HTML]{C00000} \#5 no-seed} & 0.0493      \\ \hline
{\color[HTML]{C00000} \#6 no-seed} & 0.0528      \\ \hline
\end{tabular}
\end{table*}

\begin{table*}[]
\caption{Time intervals and positions (in GSE coordinates) for all presented events corresponding to MMS and ARTEMIS measurements. \label{tab:S4}}
\begin{tabular}{|l|l|c|c|l|c|c|}
\hline
{\color[HTML]{1F497D} \textbf{}}   & \multicolumn{1}{c|}{\begin{tabular}[c]{@{}c@{}}MMS\\ (GSE)\end{tabular}} & \begin{tabular}[c]{@{}c@{}}MMS\\ (Start UTC)\end{tabular}     & \begin{tabular}[c]{@{}c@{}}MMS\\ (Stop UTC)\end{tabular}      & \multicolumn{1}{c|}{\begin{tabular}[c]{@{}c@{}}ARTEMIS\\ (GSE)\end{tabular}} & \begin{tabular}[c]{@{}c@{}}ARTEMIS\\ (Start UTC)\end{tabular} & \begin{tabular}[c]{@{}c@{}}ARTEMIS\\ (Stop UTC)\end{tabular}  \\ \hline
{\color[HTML]{1F497D} \#1 seed}    & 10.8, 11.1, 5.1                                                          & \begin{tabular}[c]{@{}c@{}}2017-12-17\\ 17:52:00\end{tabular} & \begin{tabular}[c]{@{}c@{}}2017-12-17\\ 17:54:00\end{tabular} & 65.2, -5.3, 5.1                                                              & \begin{tabular}[c]{@{}c@{}}2017-12-17\\ 17:39:00\end{tabular} & \begin{tabular}[c]{@{}c@{}}2017-12-17\\ 17:43:00\end{tabular} \\ \hline
{\color[HTML]{1F497D} \#2 seed}    & 10.2,-6.0, 5.9                                                           & \begin{tabular}[c]{@{}c@{}}2021-02-13\\ 11:05:00\end{tabular} & \begin{tabular}[c]{@{}c@{}}2021-02-13\\ 11:06:10\end{tabular} & 57.6, 18.6, -4.9                                                             & \begin{tabular}[c]{@{}c@{}}2021-02-13\\ 11:02:00\end{tabular} & \begin{tabular}[c]{@{}c@{}}2021-02-13\\ 11:10:00\end{tabular} \\ \hline
{\color[HTML]{1F497D} \#3 seed}    & 8.7, 17.0, 4.7                                                           & \begin{tabular}[c]{@{}c@{}}2018-12-10\\ 06:27:00\end{tabular} & \begin{tabular}[c]{@{}c@{}}2018-12-10\\ 06:31:00\end{tabular} & 50.1, 35.6, -0.2                                                             & \begin{tabular}[c]{@{}c@{}}2018-12-10\\ 06:16:00\end{tabular} & \begin{tabular}[c]{@{}c@{}}2018-12-10\\ 06:19:00\end{tabular} \\ \hline
{\color[HTML]{1F497D} \#4 seed}    & 16.6, 1.9, 6.3                                                           & \begin{tabular}[c]{@{}c@{}}2019-01-05\\ 17:38:00\end{tabular} & \begin{tabular}[c]{@{}c@{}}2019-01-05\\ 17:41:00\end{tabular} & 63.3, -3.5, 1.3                                                              & \begin{tabular}[c]{@{}c@{}}2019-01-05\\ 17:20:00\end{tabular} & \begin{tabular}[c]{@{}c@{}}2019-01-05\\ 17:35:00\end{tabular} \\ \hline
{\color[HTML]{1F497D} \#5 seed}    & 9.3, 17.4, 4.9                                                           & \begin{tabular}[c]{@{}c@{}}2018-12-10\\ 05:12:00\end{tabular} & \begin{tabular}[c]{@{}c@{}}2018-12-10\\ 05:25:00\end{tabular} & 50.4, 35.3, -0.1                                                             & \begin{tabular}[c]{@{}c@{}}2018-12-10\\ 05:17:00\end{tabular} & \begin{tabular}[c]{@{}c@{}}2018-12-10\\ 05:23:00\end{tabular} \\ \hline
{\color[HTML]{1F497D} \#6 seed}    & 9.6, 17.6, 5.0                                                           & \begin{tabular}[c]{@{}c@{}}2018-12-10\\ 04:40:00\end{tabular} & \begin{tabular}[c]{@{}c@{}}2018-12-10\\ 04:42:00\end{tabular} & 50.7, 35.0, -0.1                                                             & \begin{tabular}[c]{@{}c@{}}2018-12-10\\ 04:20:00\end{tabular} & \begin{tabular}[c]{@{}c@{}}2018-12-10\\ 04:37:00\end{tabular} \\ \hline
{\color[HTML]{C00000} \#1 no-seed} & 15.2, 5.6, 5.2                                                           & \begin{tabular}[c]{@{}c@{}}2018-01-12\\ 01:50:00\end{tabular} & \begin{tabular}[c]{@{}c@{}}2018-01-12\\ 01:52:30\end{tabular} & 38.0, -49.4, 5.1                                                             & \begin{tabular}[c]{@{}c@{}}2018-01-12\\ 01:35:00\end{tabular} & \begin{tabular}[c]{@{}c@{}}2018-01-12\\ 01:45:00\end{tabular} \\ \hline
{\color[HTML]{C00000} \#2 no-seed} & 8.7, 12.2, -8.1                                                          & \begin{tabular}[c]{@{}c@{}}2023-01-16\\ 08:21:00\end{tabular} & \begin{tabular}[c]{@{}c@{}}2023-01-16\\ 08:24:00\end{tabular} & 13.1, -55.9, -0.3                                                            & \begin{tabular}[c]{@{}c@{}}2023-01-16\\ 08:05:00\end{tabular} & \begin{tabular}[c]{@{}c@{}}2023-01-16\\ 08:15:00\end{tabular} \\ \hline
{\color[HTML]{C00000} \#3 no-seed} & 11.7, 12.2, -5.6                                                         & \begin{tabular}[c]{@{}c@{}}2021-01-12\\ 01:18:00\end{tabular} & \begin{tabular}[c]{@{}c@{}}2021-01-12\\ 01:21:00\end{tabular} & 58.2, -17.3, -1.1                                                            & \begin{tabular}[c]{@{}c@{}}2021-01-12\\ 00:56:00\end{tabular} & \begin{tabular}[c]{@{}c@{}}2021-01-12\\ 01:01:00\end{tabular} \\ \hline
{\color[HTML]{C00000} \#4 no-seed} & 1.6, -24.8, -1.5                                                         & \begin{tabular}[c]{@{}c@{}}2022-05-02\\ 18:23:00\end{tabular} & \begin{tabular}[c]{@{}c@{}}2022-05-02\\ 18:25:00\end{tabular} & 61.2, 22.2, 1.3                                                              & \begin{tabular}[c]{@{}c@{}}2022-05-02\\ 17:55:00\end{tabular} & \begin{tabular}[c]{@{}c@{}}2022-05-02\\ 18:03:00\end{tabular} \\ \hline
{\color[HTML]{C00000} \#5 no-seed} & -2.3, 23.4, -9.5                                                         & \begin{tabular}[c]{@{}c@{}}2022-11-24\\ 04:16:00\end{tabular} & \begin{tabular}[c]{@{}c@{}}2022-11-24\\ 04:18:00\end{tabular} & 55.1, 22.2, 1.3                                                              & \begin{tabular}[c]{@{}c@{}}2022-11-24\\ 03:48:00\end{tabular} & \begin{tabular}[c]{@{}c@{}}2022-11-24\\ 03:54:00\end{tabular} \\ \hline
{\color[HTML]{C00000} \#6 no-seed} & 14.4, 7.4, 6.1                                                           & \begin{tabular}[c]{@{}c@{}}2018-12-14\\ 04:21:00\end{tabular} & \begin{tabular}[c]{@{}c@{}}2018-12-14\\ 04:22:00\end{tabular} & 16.1,61.9,-3.8                                                               & \begin{tabular}[c]{@{}c@{}}2018-12-14\\ 04:00:00\end{tabular} & \begin{tabular}[c]{@{}c@{}}2018-12-14\\ 04:10:00\end{tabular} \\ \hline
\end{tabular}
\end{table*}
\clearpage

\end{document}